\documentclass[twocolumn]{aastex631}

\shorttitle{Star formation in NGC~4395} 
\shortauthors{Payel et al.}
\graphicspath{{./}{figures/}}
\usepackage{rotating}
\begin{document}

\title{Star formation in the dwarf Seyfert galaxy NGC~4395: Evidence for both AGN and SNe feedback?}

\correspondingauthor{Payel Nandi}
\email{payel.nandi@iiap.res.in}

\author{Payel Nandi}
\affiliation{Indian Institute of Astrophysics, Block II, Koramangala, Bangalore, India}
\affiliation{Indian Institute of Science, Bangalore, India}

\author{C. S. Stalin}
\affiliation{Indian Institute of Astrophysics, Block II, Koramangala, Bangalore, India}

\author{D. J. Saikia}
\affiliation{Inter-University Centre for Astronomy and Astrophysics (IUCAA), Post Bag 4, Ganeshkhind, Pune 411 007, India}

\author{S. Muneer}
\affiliation{Indian Institute of Astrophysics, Block II, Koramangala, Bangalore, India}

\author{George Mountrichas}
\affiliation{Instituto de Fisica de Cantabria (CSIC-Universidad de Cantabria), Avenida de los Castros, 39005 Santander, Spain}

\author{Dominika Wylezalek}
\affiliation{Astronomical Calculation Institute, Univeristy of Heidelberg, D-69120 Heidelberg, Germany}

\author{R. Sagar}
\affiliation{Indian Institute of Astrophysics, Block II, Koramangala, Bangalore, India}

\author{Markus Kissler-Patig}
\affiliation{European Space Agency (ESA), European Space Astronomy Centre (ESAC), Camino Bajo del Castillo s/n, 28692 Villanueva de la Ca$\tilde{n}$ada, Madrid, Spain}

\begin{abstract}
We present a detailed multi-wavelength study of star formation in the dwarf galaxy NGC 4395 which hosts an active galactic nucleus (AGN). From our observations with the Ultra-Violet Imaging Telescope, we have compiled a catalogue of 284 star forming (SF) regions, out of which we could detect 120 SF regions in H$\alpha$ observations. Across the entire galaxy, we found the extinction corrected star formation rate (SFR) in the far ultra-violet (FUV) to range from 2.0 $\times$ 10$^{-5}$ M$_\odot$yr$^{-1}$ to 1.5 $\times$ 10$^{-2}$  M$_\odot$yr$^{-1}$ with a median of 3.0 $\times$ 10$^{-4}$ M$_\odot$yr$^{-1}$ and age to lie in the range of $\sim$ 1 to 98 Myr with a median of 14 Myr. In H$\alpha$ we found the SFR to range from 7.2 $\times$ 10$^{-6}$ M$_\odot$yr$^{-1}$ to 2.7 $\times$ 10$^{-2}$ M$_\odot$yr$^{-1}$ with a median of 1.7 $\times$ 10$^{-4}$ M$_\odot$yr$^{-1}$ and age to lie between 3 to 6 Myr with a median of 5 Myr. The stellar ages derived from H$\alpha$ show a gradual decline with galactocentric distance. We found three SF regions close to the center of 
NGC~4395 with high SFR both from H$\alpha$ and UV which could be attributed to feedback effects from the AGN. We also found six other SF regions in one of the spiral arms having higher SFR. These are very close to supernovae remnants which could have enhanced the SFR locally. We obtained a specific SFR (SFR per unit mass) for the whole galaxy 4.64 $\times$ 10$^{-10}$ yr$^{-1}$.
\end{abstract}

\keywords{galaxies$:$ active $--$ galaxies$:$ dwarf $--$ galaxies$:$ Seyfert $--$ galaxies$:$ star formation $--$ radio continuum$:$ ISM $--$ galaxies$:$ individual$:$ NGC 4395}

\section{Introduction}
In the Seyfert category of active galactic nuclei (AGN), circumnuclear star formation is commonly observed \citep{1998MNRAS.300..388D,2015MNRAS.451.3173A,2018MNRAS.477.1086H,2019MNRAS.487.3958D}. However, works to clarify the relationships between the central AGN and star formation in their surroundings are still ongoing. There are suggestions for causal and evolutionary connections. The difficulty in establishing the connection is due to differences in the spatial and temporal scales between AGN and starburst activities. Available studies on AGN hosts probe star formation on scales of kilo-parsecs (kpc) to a few hundreds of parsecs (pc) \citep{2022MNRAS.512.3906R,2021A&A...648A..17V}. However, accretion
processes that feed the AGN occur on scales of the order of parsecs. 
The influence AGN have on their host galaxies is expected to decrease with an increase in the distance of the regions from the central AGN \citep{2015AJ....150...43T}. Therefore, to establish the connection between AGN and star formation activities, one needs to probe different spatial scales in the hosts of AGN.

The star formation rate (SFR) of a galaxy is an important parameter.
Several SFR measures are available in the literature \citep{2013seg..book..419C,2012ARA&A..50..531K}.
Observationally, the star formation nature of a galaxy can be
characterized by measuring its luminosity, which can then be used
to find the number of O, B stars. Because of their high luminosity and temperature, Ultra-Violet (UV) observations are the most straightforward
SFR indicator. Alternatively, H$\alpha$ observations in the
optical are generally used to estimate SFR \citep{1998ARA&A..36..189K}.
As UV photons, and to a lesser extent optical photons, are affected by dust attenuation, observations at other wavelengths are also used to estimate SFR.
Other estimators of SFR are via the radio and infrared (IR) emission emitted
from the star forming (SF) regions \citep{2012ARA&A..50..531K,2013seg..book..419C}.
The radio emission from such regions can be due to thermal free-free (bremsstrahlung) and non-thermal (synchrotron) processes
\citep{1992ARA&A..30..575C}. An advantage of using radio emission
to trace SFR is that it is free from dust extinction, unlike that of
the observations in the UV and optical bands that are sensitive to dust.  There are a number of studies in the literature that focus on star
formation properties of Seyfert type AGN using optical observations
\citep[e.g.][]{1998MNRAS.300..388D,2001A&A...374..932G,2018MNRAS.477.1086H}. Though observations in the UV band are an alternative tracer to investigate the star formation characteristics of
the host galaxies of AGN, such observations are
limited \citep{1997ApJ...488L..71C,1997AJ....114..575F,2009MNRAS.399..842M}.

Although star formation characteristics of galaxies hosting an AGN have been studied reasonably extensively, similar studies for dwarf galaxies with M$_{\ast}$ $<$ 10$^{10}$ M$_{\odot}$ are extremely limited. Understanding the nature
of star formation in the AGN-hosting dwarf galaxies is important as these
systems are expected to be powered by small or intermediate mass black holes \citep{2022NatAs...6...26R}. Studies of such objects in the nearby Universe may provide valuable insights in understanding how similar objects may have evolved in the early Universe. Earlier, theoretical studies on the
regulation of star formation in dwarf galaxies have been attributed to radiation
from young stars and supernova (SNe) explosions \citep{2002MNRAS.333..156B}.
However, recent theoretical studies do indicate that AGN could play a significant role in regulating star formation in dwarf galaxies \citep{2022MNRAS.516.2112K}. Observationally, there is evidence of AGN feedback operating in dwarf galaxies covering angular sizes smaller than about an arcmin \citep{2018MNRAS.476..979P}. In addition to such studies, it is also important to extend similar studies to large range of spatial scales.
Also, close correlations are known to exist between the mass of supermassive black holes that power AGN and various properties of their host galaxies \citep{1995ARA&A..33..581K,1998AJ....115.2285M,2000ApJ...539L...9F,2000ApJ...539L..13G,2009ApJ...698..198G}. The dwarf galaxies with smaller black hole masses appear to follow the
same scaling relationships as of higher mass galaxies, although with a
larger scatter.  These correlations and studies of the effects of jets also suggest that AGN can influence their galaxies through feedback process \citep{2012ARA&A..50..455F,2021A&A...648A..17V}. Though AGN feedback effects are observationally known in massive galaxies such as AGN inhibiting as well as enhancing star formation, observational evidences for AGN feedback in dwarf galaxies is very limited \citep{2018MNRAS.476..979P, 2022Natur.601..329S, 2020ARA&A..58..257G, 2020ApJ...896...10B}. Therefore, detailed studies on star formation characteristics of dwarf galaxies hosting AGN are needed, firstly, to characterize their star formation properties and secondly to find evidences of the feedback process if any in them.

NGC~4395 is a nearby dwarf galaxy situated at a distance of 4.3 Mpc \citep{2004AJ....127.2322T}. It hosts a low luminosity Seyfert type AGN and is powered by a black hole with mass between 10$^4$ $-$ 10$^5$ M$_{\odot}$. For example, from reverberation mapping
observations in the UV band carried out with the Space Telescope Imaging Spectrograph on the
Hubble Space Telescope, \cite{2005ApJ...632..799P} found the mass of the black hole in NGC~4395 as M$_{BH}$ = (3.6 $\pm$ 1.1) $\times$ 10$^{5}$ M$_{\odot}$. However,
recently, using the response of the H$\alpha$ emission line to the
variations in the V-band continuum, \cite{2019NatAs...3..755W} found
a lower value of M$_{BH}$ = (9.1$^{+1.5}_{-1.6}$ $\times$ 10$^3$) M$_{\odot}$.
NGC~4395 shows other characteristics of Seyfert 1 galaxies, such as
variable emission in the optical and X-rays \citep{1999MNRAS.305..109L}
and a collimated radio structure with components on opposite sides of the galaxy nucleus \citep{2001ApJ...553L..23W,2018A&A...616A.152S,2022MNRAS.514.6215Y}.
However, it has been found
to have a low bolometric luminosity between 10$^{40}$ $-$ 10$^{41}$ erg s$^{-1}$
\citep{1999MNRAS.305..109L,2003ApJ...588L..13F,2019MNRAS.486..691B}.

\cite{2020AstBu..75..234S} have identified about a hundred SF regions in NGC~4395 from H$\alpha$ observations with an angular resolution of 2.9 arcsec, and have made a limited study of comparing their infrared fluxes and UV surface brightness using archival GALEX data. However, for a detailed understanding of SF in this galaxy, deeper and higher resolution observations in the UV and a multi-wavelength approach, including radio, are required.
We, therefore, carried out a detailed investigation of star formation in NGC~4395 using our UV observations on scales of few hundreds of pc to few kpc. In addition to UV, we also observed NGC~4395 in H$\alpha$, and analysed the nature of those SF regions identified from UV using multi-wavelength data. In Section 2, we describe 
the observations and the data reduction procedures. Section 3 describes the 
identification of the SF regions. The properties of the detected SF regions, along with the global star formation properties of the galaxy are presented in Section 4. A discussion of the results are presented in Section 5, which is followed by a summary in the final Section.
 In this work, we 
consider a flat $\Lambda$CDM cosmology with 
$H_0$=70 km$^{-1}$ s$^{-1}$ Mpc$^{-1}$, $\Omega_{\Lambda}$ = 0.7 and $\Omega_{m}$=0.3.

\begin{table}
\caption{Details of the galaxy NGC~4395}
\centering
\begin{tabular}{p{0.2\linewidth} p{0.3\linewidth} p{0.25\linewidth}} 
\hline
     Parameter          & Value          & Reference \\ 
     \hline
     RA(J2000)          & 12:25:48.860   & \citealt{2022arXiv220800211G}  \\
     DEC(J2000)         & +33:32:48.711  & \citealt{2022arXiv220800211G}  \\
     z                  & 0.00106        &  \citealt{1998AJ....115...62H} \\ 
     Morphology         &  SA(s)m        & \citealt{1991rc3..book.....D}  \\
     Semi-major axis    &  6.59 arcmin   & \citealt{1991rc3..book.....D}  \\
     Semi-minor axis    &  5.48 arcmin   & \citealt{1991rc3..book.....D}  \\
     Position angle     &  147 deg       & \citealt{1991rc3..book.....D}  \\
     Type              &  Seyfert 1      & \citealt{1989ApJ...342L..11F} \\
     Stellar mass      & 2.5$\times$10$^9$ M$_\odot$  &   \citealt{2020AstBu..75..234S}  \\
     Distance          & 4.3 Mpc         & \citealt{2004AJ....127.2322T}\\
     Inclination angle & 38 $\deg$       & \citealt{1989ApJ...342L..11F}\\
    \hline
\end{tabular}
\label{table-1}
\end{table}

\begin{figure*}
    \includegraphics[scale=1.2]{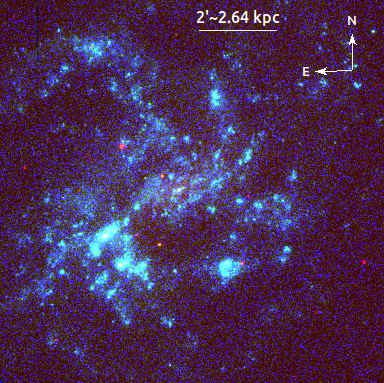}
    \caption{The colour composite image of 
     NGC~4395 covering a region
of 10 $\times$ 10 square arcmin. Here, red is the R-band image from HCT, 
green is the NUV image in the N263M 
filter and blue is the FUV image in F148W
filter. The filter details are in Table 2. }
    \label{figure-1}
\end{figure*}

\section{Observations and data reduction}
For this work, we used our new observations (UV and H$\alpha$) as well as archival data   (radio, IR, and broad band optical).

\subsection{Ultra-Violet}
We observed NGC~4395 in the UV 
band using the Ultra-Violet Imaging Telescope (UVIT; \citealt{2017AJ....154..128T}), one 
of the payloads onboard 
{\it AstroSat} India's first multi-wavelength astronomical 
observatory \citep{2006AdSpR..38.2989A} 
launched by the Indian Space Research Organization
(ISRO) in September 2015. The details of the source are given in Table \ref{table-1}. 
UVIT observes simultaneously in three channels, namely
far-UV (FUV; 1300$-$1800 \AA), near-UV, (NUV; 2000$-$3000 \AA) and the visible 
(VIS; 3200$-$5500 \AA) over a field-of-view of 28$^{\prime}$ diameter with 
spatial resolution better than 1.5 arcsec in UV. Details of UVIT can be found in 
\cite{2020AJ....159..158T}. 

NGC~4395 was observed by UVIT on 27 February 2018 in the photon counting mode with the default
frame rate of $\sim$34 frames per second. We directly downloaded the level2 (L2) science ready 
images processed using the official L2 pipeline \citep{2022arXiv220307693G} by 
the UVIT-Payload Operations Centre at the Indian Institute of Astrophysics, 
Bangalore and transferred to the Indian Space Science Data Center (ISSDC) for archival 
and dissemination. In the combined images from ISSDC, we found the exposure 
time to be lesser than the sum of the individual exposures (orbit wise images). We therefore,
based our analysis from the reduced individual L2 images. We firstly aligned 
the individual images using the Image Reduction and Analysis Facility (IRAF; \citealt{1986SPIE..627..733T}) and combined those aligned individual images to create the 
combined image filter-wise. Astrometry on the combined images was carried out 
using stars available on the image frames with their ($\alpha$, $\delta$) 
positions taken from {\it Gaia-DR3} \citep{2022arXiv220800211G} through 
custom developed {\it Python} scripts. These astrometrically
corrected combined image frames were used for further analyses. 

\subsection{Optical}
Optical observations in the narrow H${\alpha}$ and broad R-band were obtained 
on 22 December 2021, using the Himalayan Faint Object Spectrograph Camera (HFOSC)  instrument mounted on the 
2m Himalayan Chandra Telescope (HCT) situated in the Indian Astronomical Observatory, 
Hanle, India \citep{2014PINSA..80..887P}. HFOSC is equipped with a CCD with a 
dimension of 2048 $\times$ 4096 pixels. The detector has two amplifiers, 
A and B which can be operated in both high and low gain modes. The readout 
noise for amplifier A in high and low gain modes are 4.8 electrons  and 
8.0 electrons respectively, with gain values of 1.22 electrons/ADU and 
5.6 electrons/ADU. Readout noise and gain for amplifier B in high and low 
gain modes are 5.1 electrons, 8.0 electrons and 1.21 electrons/ADU, 5.6 
electrons/ADU.  For imaging, we used amplifier A in high gain mode of 
1.22 electrons/ADU.  We used only the central 2048 $\times$ 2048  pixels 
covering a field of view of 10 $\times$ 10 square arcmin with an image scale of 
0.296$^{\prime\prime}$/pixel. Three exposures each of 600 seconds or a total of 1800 sec was taken 
in H${\alpha}$ narrow band and one frame of 300 seconds in the R-band. These observations have spatial resolution between 1.9$-$2.2 arcsec.
The log of observations is given in 
Table \ref{table-2}.  
Standard procedures were followed to reduce the acquired data that includes 
bias subtraction, flat fielding and cosmic ray removal using  
IRAF software \citep{1986SPIE..627..733T}.
The three frames in H${\alpha}$ were then combined and astrometrically 
calibrated using {\it Gaia-DR3} \citep{2022arXiv220800211G}  catalogue.
 
For flux calibration in the narrow band H$\alpha$ we 
also observed one standard spectrophotometric star Feige 66. We convolved the 
observed spectrum  of Feige 66 from \cite{1990AJ.....99.1621O} with the H${\alpha}$ filter response and estimated the  magnitude. The derived magnitude 
in H${\alpha}$ observation was converted
to standard magnitude using the difference between the instrumental and standard
magnitude of Feige 66. To get the standard magnitude in R-band, we used the 
technique of differential photometry in conjunction with the relations from 
SDSS\footnote{http://sdss.org}. The transformation equations from SDSS were used to get the standard R-band magnitude of few stars in the acquired R-band image based on their SDSS u, g, r, i, z magnitudes. We then subtracted the  
calibrated R-band image from the calibrated H$\alpha$ narrow band image to get 
the calibrated H$\alpha$ line image as \citep{1990PASP..102.1217W}
\begin{equation}
f(line) = \frac{[f_{\lambda}(N) - f_{\lambda}(B)]W(N)} {[1-W(N)/W(B)]}
\end{equation}
Here, $f_{\lambda}(N)$ and $f_{\lambda}(B)$ are the flux densities
in the narrow H$\alpha$ and broad R-band, while W(N) and W(B)
are the widths of the narrow H$\alpha$ and broad R-band filters respectively. 
The composite image of NGC~4395 from R, NUV, and FUV filters 
covering the central 10 $\times$ 10 square arcmin region is shown in 
Fig.~\ref{figure-1}.

In addition to the observations acquired in H$\alpha$, we also used
archival data in broad band U, B, V, R, and I filters from the Nordic
Optical Telescope (NOT; \citealt{https://doi.org/10.26132/ned1}). We carried out astrometry
as well as photometric calibration on these archival images using the technique of differential photometry.

\subsection{Infrared}
In IR, we used archival images \citep{https://doi.org/10.26132/ned1} in the J, H, and K$_s$ bands
from 2MASS \citep{2006AJ....131.1163S}, W1 (3.4 $\mu$m), W2 (4.6 $\mu$m), and W3 (12 $\mu$m) bands \citep{https://doi.org/10.26132/ned1}
 taken  from the {\it Wide-field Infrared Survey Explorer}
(WISE; \citealt{2010AJ....140.1868W}). Also, in the far-IR, we used
data at  24 $\mu$m, 70 $\mu$m, and 160 $\mu$m \citep{https://doi.org/10.26132/ned1} from the 
MIPS (Multiband Imaging Photometer for
Spitzer)  instrument \citep{2004ApJS..154...25R} in Spitzer. The
measurements obtained at these wavelengths were converted to flux units
using known scaling factors. For example, in the case of the 
images from 2MASS, to convert from instrumental measurements to 
flux units we used the factors given in the respective image headers.
For {\it WISE} we used the factors available online\footnote{https://wise2.ipac.caltech.edu/docs/release/allsky/expsup/sec2\_3f.html}. Similarly for MIPS, we followed the 
procedure given in the MIPS instrument handbook.\footnote{https://irsa.ipac.caltech.edu/data/SPITZER/docs/mips/mipsinstrumenthandbook/41/}

\subsection{Radio}
We used observations from the Very Large Array (VLA) carried out at 1.4 GHz 
in its C-configuration. We also have archival data at 1.4 GHz from the VLA 
in A and B configurations. However, in the A and B configuration data, 
only the central AGN is detected. Therefore, in this work we used
only the data in C configuration, the processed image of which was taken from the
archives of VLA.\footnote{www.vla.nrao.edu/astro/nvas/}. We also used 21cm HI emission line data from the Westerbork Synthesis Radio Telescope (WSRT; \citealt{2008ASPC..396..267H}).

\begin{figure*}
 \hspace*{-0.5cm}\includegraphics[scale=0.9]{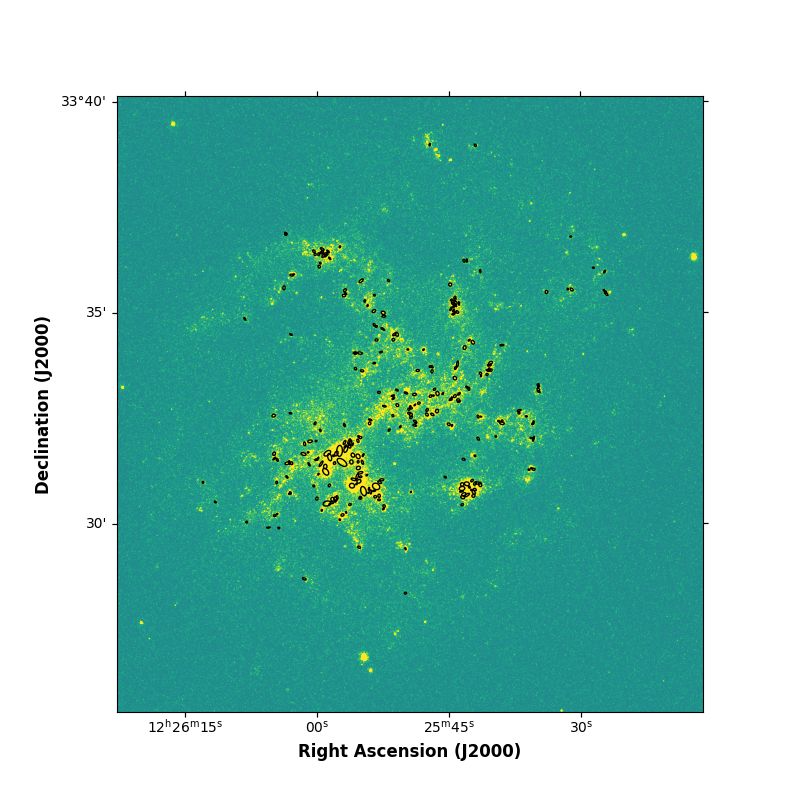}
    \caption{The image of NGC~4395 in the UV F148W filter. The regions
that have been identified are marked.}
    \label{figure-2}
\end{figure*}

\begin{table*}
\centering
\caption{Log of observations}
\begin{tabular}{cccccc}
\hline
Filter name & Central wavelength &Bandwidth & Telescope & Date of observation &  Exposure time \\
        & (\AA) & (\AA) &  &  &(sec) \\
      \hline
      F148W     & 1481 & 50.0 & UVIT &  27/02/2018  & 1348 \\
      F172M     & 1717 & 12.5 & UVIT &  27/02/2018  & 6348 \\
      N219M     & 2196 & 27.0 & UVIT &  27/02/2018  & 6386 \\
      N263M     & 2632 & 27.5 & UVIT &  27/02/2018  & 1355 \\
      656.3(10) & 6563 & 10   & HCT  &  22/12/2021  & 1800 \\
      R         & 6020 & 1000 & HCT  &  22/12/2021  & 300  \\ 
      \hline
\end{tabular}
\label{table-2}
\end{table*}

\section{Analysis}
UVIT observations were carried out in four filters (see Table \ref{table-2}). 
For analysis, we explicitly used data from all the filters. To calculate  
SFR and age we used F148W and N263M filters as these have relatively larger
effective area (indicative of the overall sensitivity of the photometric system) compared with other filters. However, for the calculation of color excess, E(B-V) we used one FUV filter (F172M) and one NUV filter (N219M) that are
closely spaced in wavelength.

\subsection{Identification of star forming regions}
To identify SF regions in NGC~4395, we used the final combined image taken in F148W. We used the Source Extractor Software 
(SExtractor; \citealt{1996A&AS..117..393B})
with  the following parameters, namely DETECT\_THRESH = 5 $\sigma$,  
DETECT\_MINAREA = 11 and DEBLEND\_THRESHOLD = 32. Using the above criteria, 
we identified
a total of 284 SF regions after removal of some overlapping regions. 
The identified SF regions marked on the F148W filter image are shown in Fig.~\ref{figure-2}. The same criteria were also used
to identify SF regions in the narrow band H$\alpha$ filter.  
Out of the 284 regions detected in FUV, we could detect 120 regions that 
have sufficient signal to noise ratio in the H$\alpha$ continuum subtracted image.

\subsection{Size of the SF regions}
\begin{figure}
    \centering
    \includegraphics[scale=0.36]{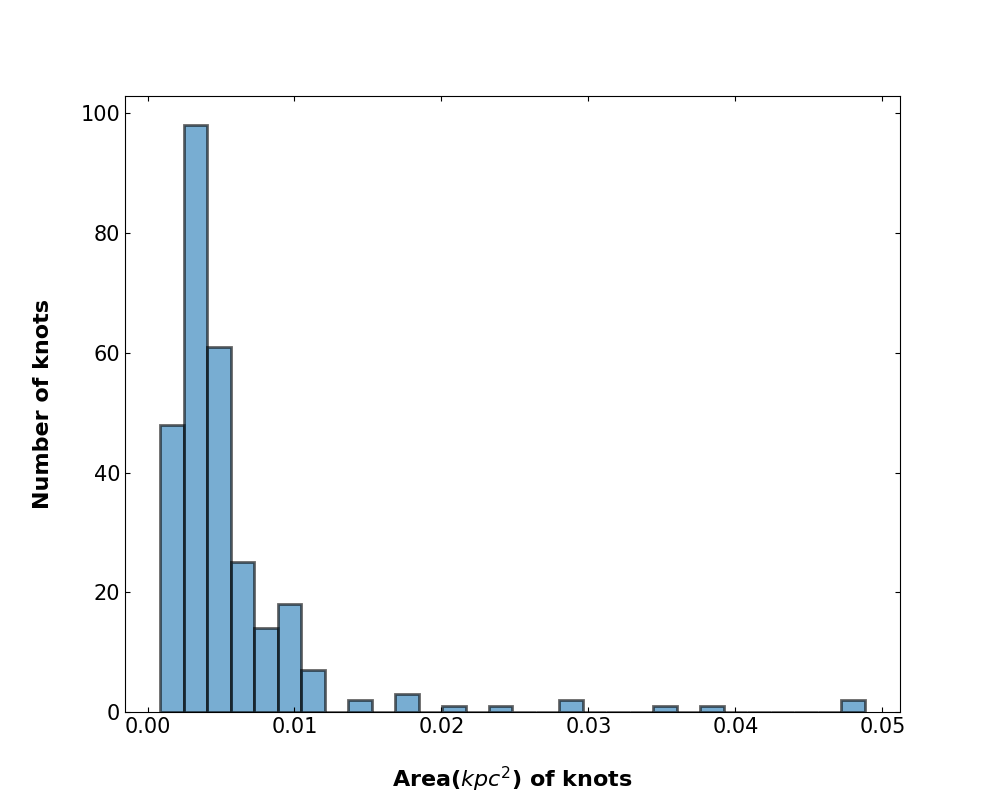}
    \caption{Distribution of the area of the SF regions identified in FUV.}
    \label{fig:area}
\end{figure}

After detecting SF regions, we checked their sizes. SExtractor gives us the semi-major and semi-minor length in pixel coordinate and position angle in degree corresponding to each SF region. We converted these to arcsec and then to pc unit. At the redshift z = 0.00106 of NGC~4395, one arcsec corresponds to 22 pc. 
The measured sizes of the SF regions in UV and H$\alpha$ were corrected for their respective instrumental resolutions by assuming elliptical light distribution within the apertures. We could correct for instrumental resolution for 225 regions out of 284 detected regions in UV and 77 regions out of 120 regions. For the remaining regions with sizes similar to the instrumental resolution, we considered the apertures given by SExractor.
We noticed that the area of the SF regions, detected in UV varies from $8.6 \times 10^{-4}$ kpc$^2$ to $488.1 \times 10^{-4}$ kpc$^2$ with a median of $40.2 \times 10^{-4}$ kpc$^{2}$.
The distribution of the area of the SF regions is shown in Fig.~\ref{fig:area}.

\subsection{Photometry}
Once the SF regions were identified and the sizes were corrected for instrumental resolution in F148W and H$\alpha$ using the criteria given in 
Section 3.1 and 3.2, we carried out photometry of the regions using aperture photometry technique in 
{\it PhotUtils} \citep{larry_bradley_2020_4044744}. 
As the images in F148W, F172M, N219M, and N263M  were astrometrically aligned, 
photometry of all 284 regions was also carried out on the images acquired 
in all four filters. Similarly, photometry was also carried out in the 120 
regions identified in the continuum subtracted H$\alpha$ image. One of the 
crucial parameters for photometry is the proper subtraction 
of the background. For the determination of the background, we estimated the median count rate in randomly placed square boxes of sizes 10 $\times$ 10 pixels 
in 50 source free regions.  The mean of these 50 median values 
was taken as the background per pixel. This  background per pixel was 
subtracted from 
each of the pixels in the aperture used for photometry. The background 
subtracted count rates from each of the identified SF regions in F148W, F172M, N219M, and N263M were converted to magnitudes using the calibration 
given in \cite{2020AJ....159..158T}. Similarly, for the H${\alpha}$ 
observations the derived instrumental magnitudes of the SF
regions were converted to standard magnitudes using the procedure outlined in 
Section 2.2. We note that though the aperture used for the photometry of the SF regions do not overlap, there could be contamination from adjacent SF regions.

\subsection{Extinction correction}
The magnitudes derived for the SF regions are affected by both 
internal and Galactic extinction. To estimate the extinction in
UV and H$\alpha$ for our Galaxy, we used the relation given below 
\citep{1989ApJ...345..245C}
\begin{equation}
    \langle A(\lambda)/A(V)\rangle = a(x) + b(x)/R_v
    \label{eq:ext_mw}
\end{equation}
Here, A(V)  = 0.047 mag, is the Galactic extinction in the V-band taken from 
NED\footnote{http://ned.ipac.caltech.edu}, and $x$ is the wave number. 
The values
of $a(x)$ and $b(x)$ were evaluated following \cite{1989ApJ...345..245C}. A similar
procedure was also followed to correct for Galactic extinction in
the H${\alpha}$ image.

\begin{sidewaystable}

\centering
\caption{Catalogue of SF regions in NGC~4395. Only the first 
ten entries are shown. Here, column 1 represents the index number of the detected regions. 
 Columns 2, 3, 4, 5, and 6 are the right ascension, declination, semi-major axis, semi-minor axes and position angles of the SF regions respectively. Columns 7 and 8 are the extinction corrected magnitudes in F148W and N263M filters.  Columns 9 and 10 represent the extinction in F148W and N263M filters. Columns 11 and 12 are the extinction corrected surface density of star formation rate in F148W and N263M filters. And column 13 represents the age calculated in UV. The Table in full is available in the electronic version of the article}
\begin{tabular}{ccccccccccccc} \\ \hline
 
No & RA(2000)  & DEC(2000)& a     &  b     & PA     & mag$_{FUV}$     & mag$_{NUV}$    & A$_{FUV}$  & A$_{NUV}$ & \multicolumn{2}{c} {$\Sigma_{SFR}$(M$_{\odot}$yr$^{-1}$kpc$^{-2}$)}  & Age$_{UV}$  \\ 
  & (deg)     & (deg)    &  (arcsec)       & (arcsec)       & (deg)  & (mag)   &  (mag)  & (mag)      & (mag)     & FUV & NUV & (Myr)    \\ \hline
1  & 186.548202  & 33.508548  & 1.61 & 1.08 & -41.0 & 18.55  & 18.48  & 2.15 & 1.56 & 0.0895 & 0.0996 & 44   \\
2  & 186.523056  & 33.498431  & 2.55 & 0.86 & 5.6 & 19.88  & 19.79  & 0.66 & 0.48 & 0.0209 & 0.0237 & 47 \\
3  & 186.520573  & 33.542668  & 2.59 & 1.97 & 22.6 & 18.92  & 19.05  & 0.18 & 0.13 & 0.0215 & 0.0199  & 12 \\
4  & 186.520348  & 33.527611  & 2.46 & 1.84 & 63.5 & 18.10 & 18.35  & 1.59 & 1.15 & 0.05183 & 0.04316 &  7  \\
5  & 186.520043  & 33.503160  & 1.86 & 1.28 & 4.2 & 19.31 & 19.18 & 1.08 & 0.79 & 0.03217 & 0.03795 &  55 \\
6  & 186.518845  & 33.525167  & 2.13 & 1.22 & -61.3 & 18.92 & 18.94 & 1.05 & 0.76 & 0.04264 & 0.04354 &  28  \\
7  & 186.518100  & 33.498277  & 1.57 & 1.05 & -13.1 & 18.19  & 18.47  & 2.12 & 1.54 & 0.1316 & 0.1062 &  5  \\
8  & 186.515629  & 33.593210  & 2.72 & 1.65 & 86.4 & 18.47 & 18.38  & 1.57 & 1.14 & 0.03743 & 0.04256 &  48  \\
9  & 186.514743  & 33.614744  & 2.01 & 1.20 & -32.0 & 18.92 & 19.02  & 0.76 & 0.55 & 0.04560 & 0.04367 & 19 \\
10 & 186.514135  & 33.523861  & 3.35 & 1.17 & -47.7 & 15.62  & 18.44  & 0.61 & 0.44 & 0.02881 & 0.03465  & 60  \\ 
\hline

\label{table-3}
\end{tabular}
\end{sidewaystable}

\begin{figure*}
\vbox{
\hbox{
     \hspace*{0.5cm}\includegraphics[scale=0.32]{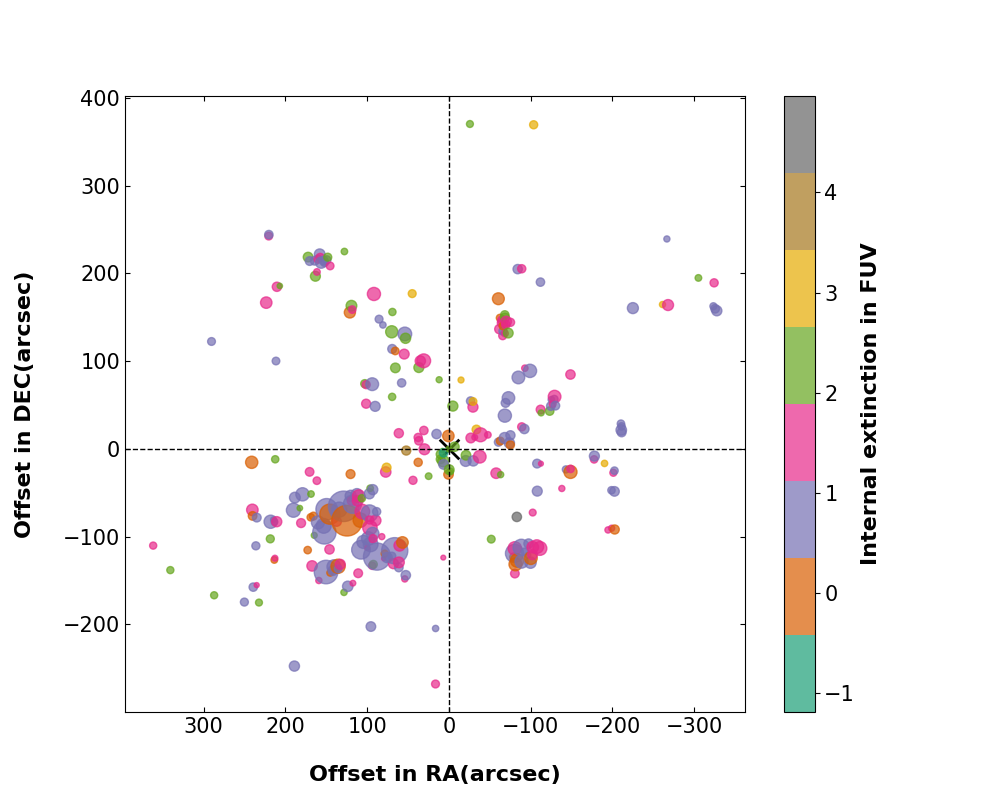}
     \hspace*{0.5cm}\includegraphics[scale=0.32]{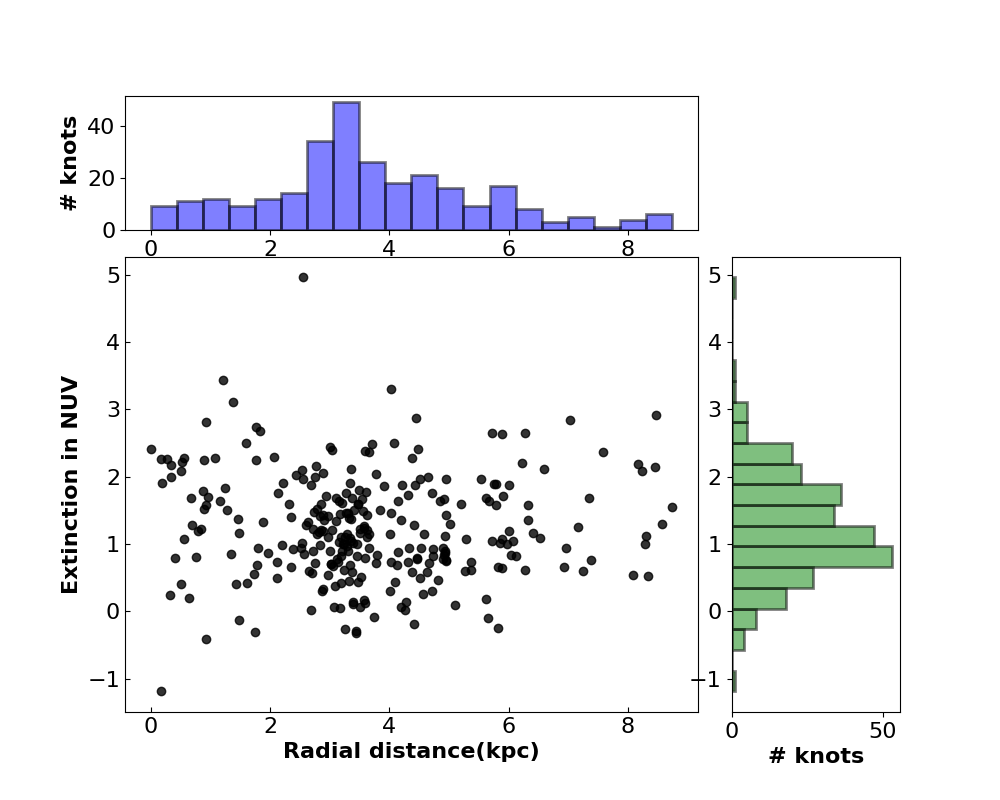}
     }

\hbox{     
      \hspace*{0.5cm}\includegraphics[scale=0.32]{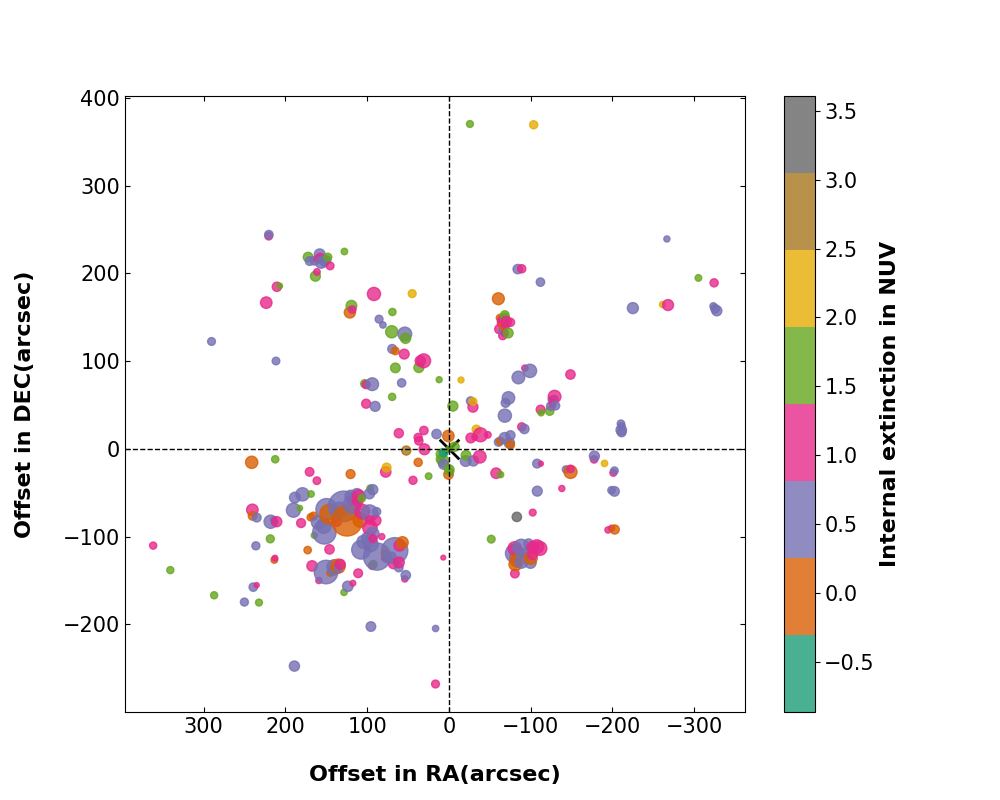}
     \hspace*{.5cm}\includegraphics[scale=0.32]{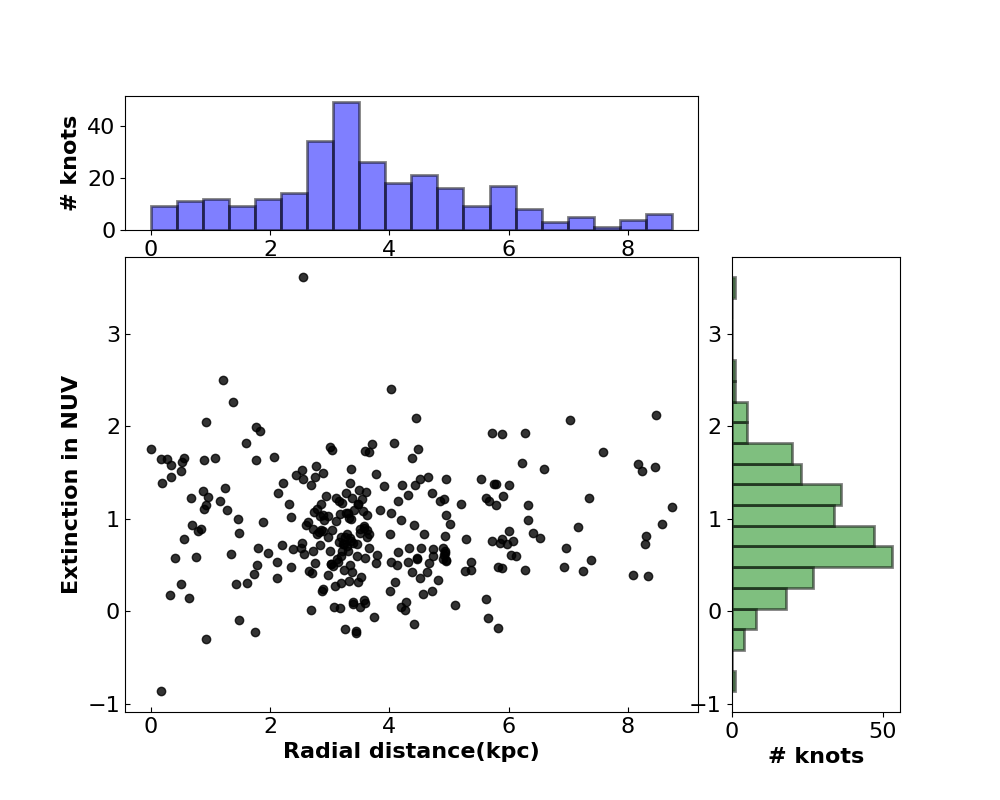}
     }
}
    \caption{The distribution of internal extinction as well as its spatial variation in the F148W (top panel) and N263M (bottom panel) filters. In the left panels, different sizes of the points correspond to the size of the SF regions.
     }
    \label{figure-3}
\end{figure*}

Dust present within each of the detected SF regions scatters
and/or absorbs the UV photons coming from those regions, making 
it difficult to estimate the absolute flux emitted from those
regions.  To correct for this in the UV band, we used the method based on the UV
spectral  slope $\beta$  defined as the slope of the power-law
function ($f_{\lambda} \propto \lambda^{\beta}$) followed by the continuum
emission of galaxies over the wavelength range of 1300$-$2600 \AA  
\citep{1994ApJ...429..582C}.  As this slope can serve as a good diagnostic for 
internal dust extinction, we adopted this to calculate the internal 
dust extinction in each of the SF regions.
For the FUV and NUV wavelengths used in this work, we calculated $\beta$
using the following relation 
\begin{equation}
\beta =  \frac{m_{FUV} - m_{NUV}} {-2.5 log(\lambda_{FUV}/\lambda_{NUV})} - 2.0
\label{eq:beta}
\end{equation}
Here m$_{FUV}$ and m$_{NUV}$ are the MilkyWay extinction corrected magnitudes in F172M and F219M filters respectively and $\lambda_{FUV}$, $\lambda_{NUV}$ are the central wavelengths of F172M and F263M filters as tabulated in Table \ref{table-2}. Using the slope obtained in Equation \ref{eq:beta}, we calculated the colour 
excess $E(B-V)$ in the SF regions as \citep{2018ApJ...853...56R}
\begin{equation}
\beta = -2.616 + 4.684 E(B-V)
\end{equation}
From $E(B-V)$, we estimated the extinction $A_{\lambda}$ as  \citep{2000ApJ...533..682C}
\begin{equation}
A_{\lambda} = k^{\prime}({\lambda}) E_s(B-V)
\end{equation}
where $E_s(B-V)= (0.44\pm 0.03)E(B-V)$
and $k^{\prime}(\lambda)$ is defined as 
\begin{equation}
k^\prime(\lambda) = 2.659(-2.15 + 1.509/\lambda - 0.198/\lambda^2 + 0.011/\lambda^3) + R^\prime_V\\
\label{eq:ext_int_2}
\end{equation}
Here, $R_V^{\prime}$ is 4.05 \citep{2000ApJ...533..682C}. For the filters 
F148W and N263M we obtained $k^{\prime}({\lambda})$ of 10.411 and 7.568 respectively.  The distribution of the internal extinction in F148W 
and N263M filters  as well as their spatial variation across NGC~4395 is 
shown in Fig.~\ref{figure-3}. 

To correct for internal extinction in H$\alpha$ we adopted the following. We calculated $\beta$ and E(B$-$V) from F172M and F219M filter images for these 120 regions, following Equations 3 and 4. We then evaluated $A_{\lambda}$ for
H$\alpha$ with the $k_{\lambda}^{\prime}$ determined as follows
\citep{2000ApJ...533..682C}
\begin{equation}
k^\prime(\lambda) = 2.659(-1.857 + 1.040/\lambda)  + R^\prime_V\\
\end{equation} 
We obtained a value of $k^\prime(\lambda)$ of 3.329. Using this we 
derived $A_{\lambda}$ for all the 120 SF regions in H$\alpha$.

\section{Properties of the Star forming regions}
\subsection{Star formation rate}
\begin{figure*}
\vbox{
\hbox{
    \includegraphics[scale=0.32]{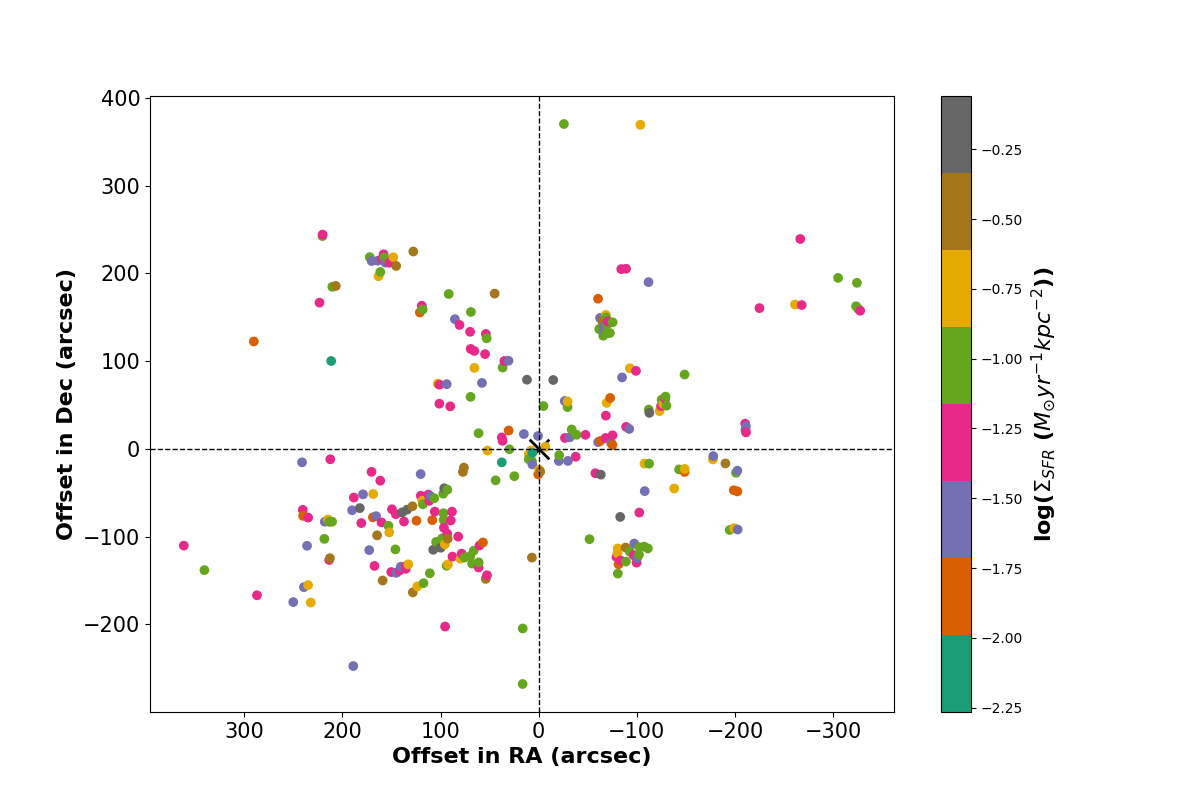}
    \hspace{-0.5cm}\includegraphics[scale=0.32]{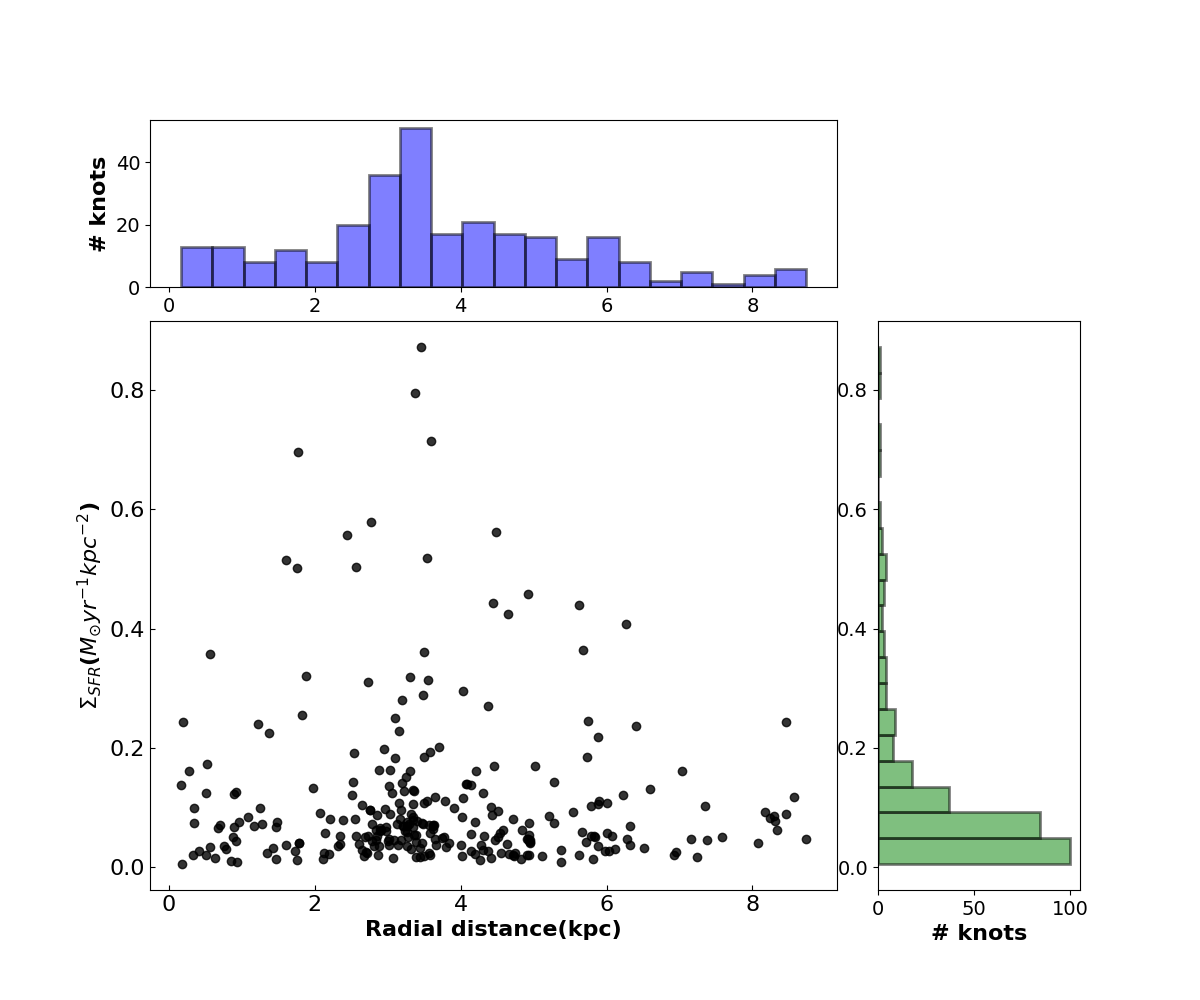}
     }

\hbox{
    \includegraphics[scale=0.32]{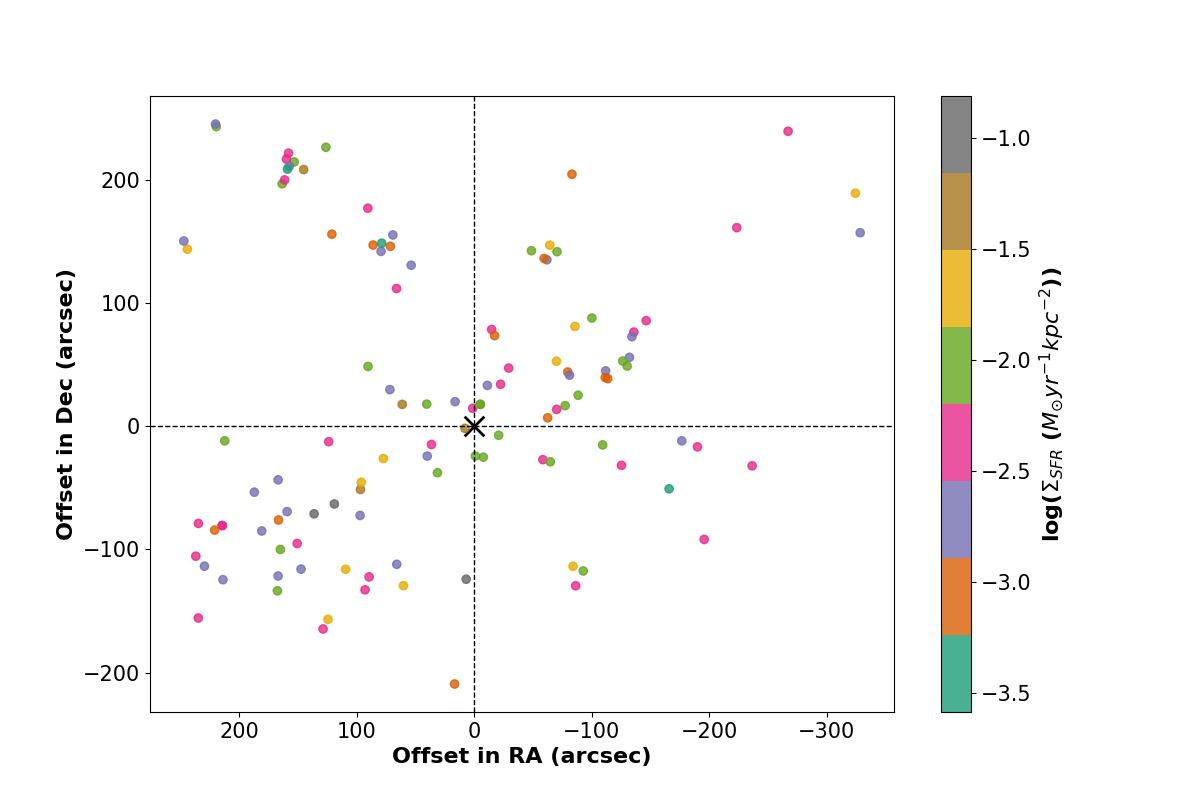}
    \hspace{-0.5cm}\includegraphics[scale=0.32]{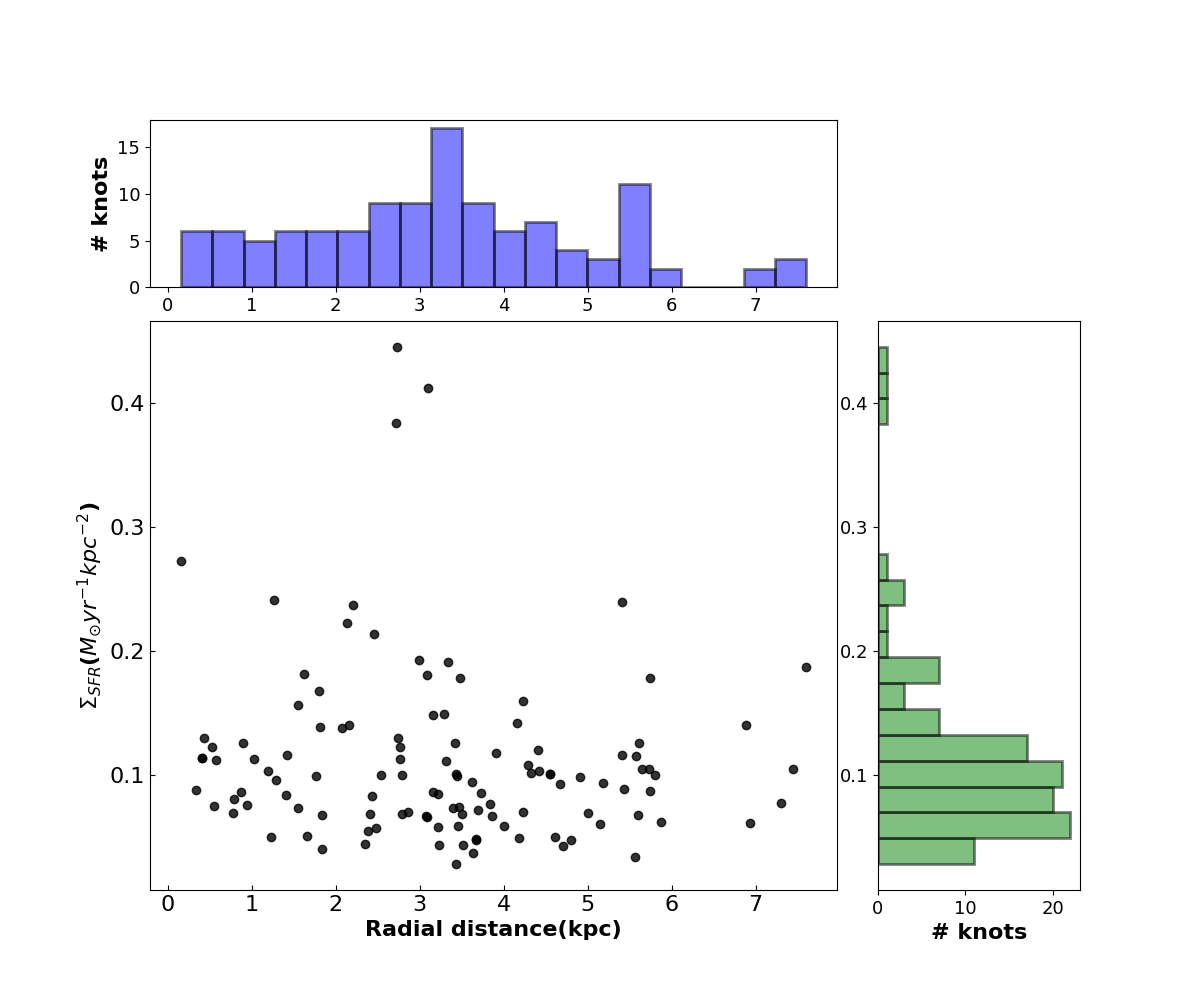}
     }
}
\caption{Distribution of $\Sigma_{SFR}$ and its spatial variation in F148W (top 
panel) and H$\alpha$ (bottom panel). Here, the central AGN is marked as a cross 
in the left panels.
}
\label{figure-4}
\end{figure*}

\begin{figure}
    \centering
    \includegraphics[scale=0.4]{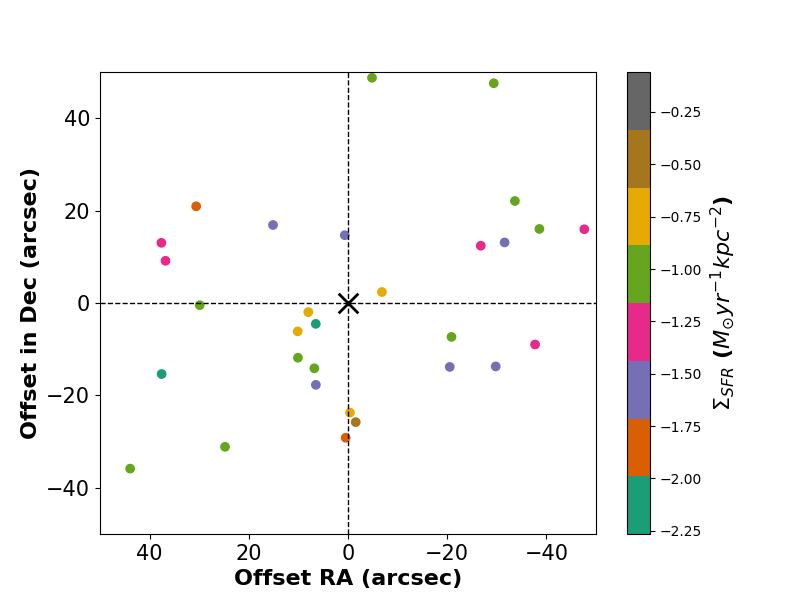}
    \caption{The variation of $\Sigma_{SFR}$ in the central 100 $\times$ 100  square arcsec$^2$ region 
in UV. The central AGN is marked as a cross in the figure. }
    \label{figure-5}
\end{figure}

The SFR provides important information on the assembly of stellar 
mass.  As different populations of stars emit efficiently at different 
wavelengths, one needs to probe the SF regions in different bands. 
In particular, UV traces the emission from young high mass stars of O, B 
type \citep{2016MNRAS.461..458D}, thereby providing best estimates of recent 
star formation over 100 Myr. Similarly, H${\alpha}$ traces star formation 
of  $<$ 10 Myr \citep{1998ARA&A..36..189K,2012ARA&A..50..531K,2013seg..book..419C}. Nearly half of the SF regions
identified in UV have also been detected in our observations in  H$\alpha$.  
We estimated the SFR in UV using F148W filter  using the following relation \citep{2007ApJS..173..267S}.

\begin{equation}
log(SFR_{FUV}(M_\odot yr^{-1}))= log[L_{FUV}(W Hz^{-1})] -21.16
\label{eq:sfr_uv}
\end{equation}
This relation (Equation \ref{eq:sfr_uv})  traces 0.1M$_\odot$ to 100M$_\odot$ stellar population having 
Chabrier initial mass function (IMF). 
We derived SFR in H${\alpha}$  following the \cite{1998ARA&A..36..189K} relation
\begin{equation}
SFR(M_\odot yr^{-1}) = 7.9 \times 10^{-42} L_{H\alpha} (erg.s^{-1})
\end{equation}
Here, $L_{FUV}$ is the intrinsic luminosity of the SF regions
in FUV and $L_{H\alpha}$ is the luminosity of the 
SF regions in H$\alpha$.  The distributions of SFR surface 
density ($\Sigma_{SFR} = SFR/Area$) and their 
variation across the galaxy in F148W and H$\alpha$ are given 
in Fig.~\ref{figure-4}.  The SFR calculated in FUV filter spans a wide 
range from 2.0 $\times$ 10$^{-5}$ M$_\odot$yr$^{-1}$ to 1.5 $\times$ 10$^{-2}$ 
M$_\odot$yr$^{-1}$ with a median value of 3.0 $\times$ 10$^{-4}$ M$_\odot$yr$^{-1}$. 
In H$\alpha$ we found the SFR to range from 7.2 $\times$ 10$^{-6}$ 
M$_\odot$yr$^{-1}$ to 2.7 $\times 10^{-2}$ M$_\odot$yr$^{-1}$ with a median 
value of 1.7 $\times 10^{-4}$ M$_\odot$yr$^{-1}$. The $\Sigma_{SFR}$ of the 
central 100 $\times$ 100 square arcsec region in F148W filter is shown in Fig.~\ref{figure-5}.

\subsection{Age of the SF regions}
We calculated the age of the SF regions using both UV colour 
(F148W $-$ N263M) and H$\alpha$ equivalent width. 

\subsubsection{UV colour}
\begin{figure}
    \centering
    \includegraphics[scale=0.33]{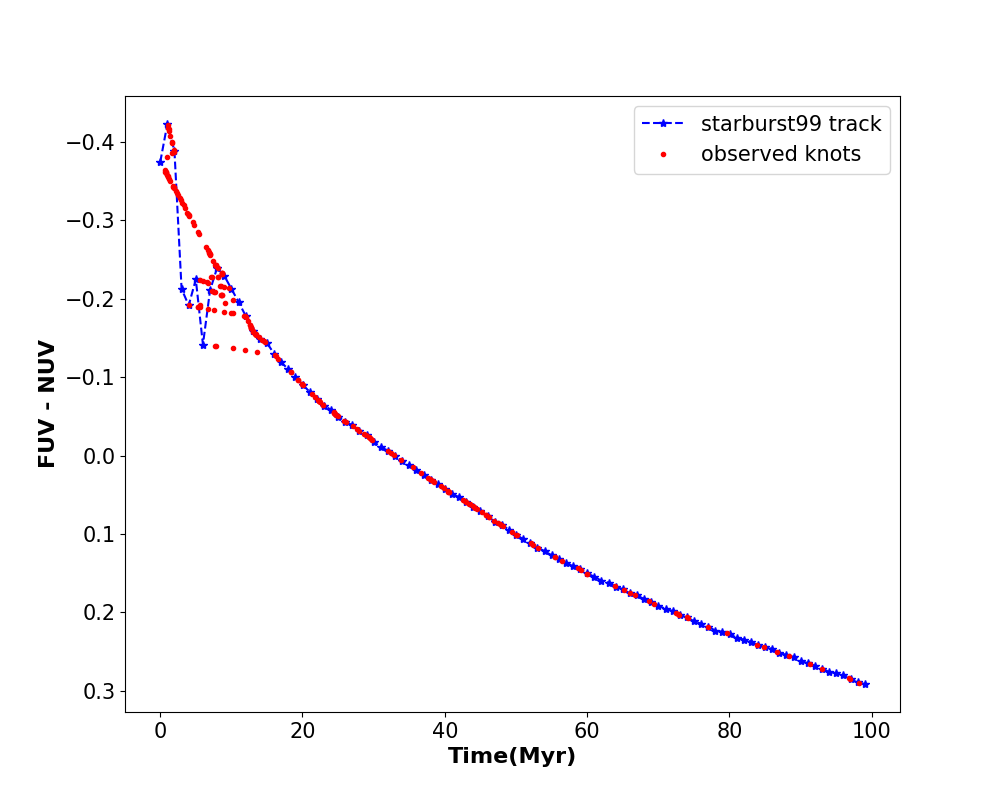}
    \caption{Extinction corrected observed (FUV$-$NUV) color (red circles) 
against age. The blue dashed line is the synthetic track derived from 
Starburst99 model.}
    \label{fig:figure-6}
\end{figure}

\begin{figure*}
\vbox{
\hbox{
     \includegraphics[scale=0.33]{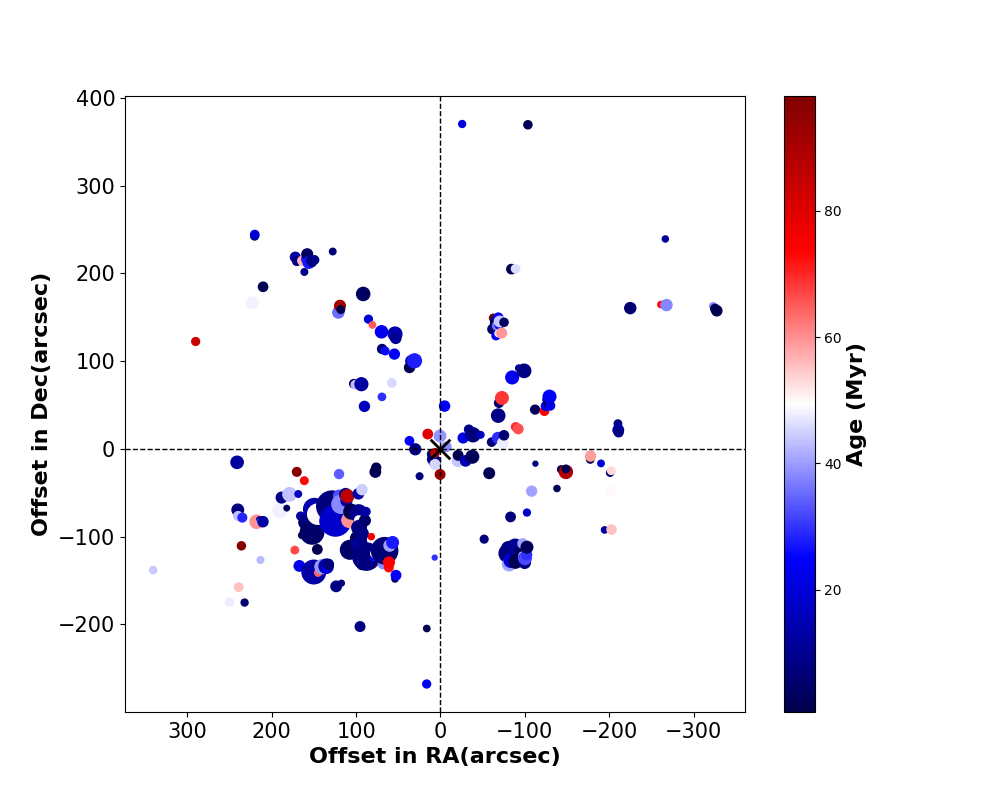}
     \includegraphics[scale=0.33]{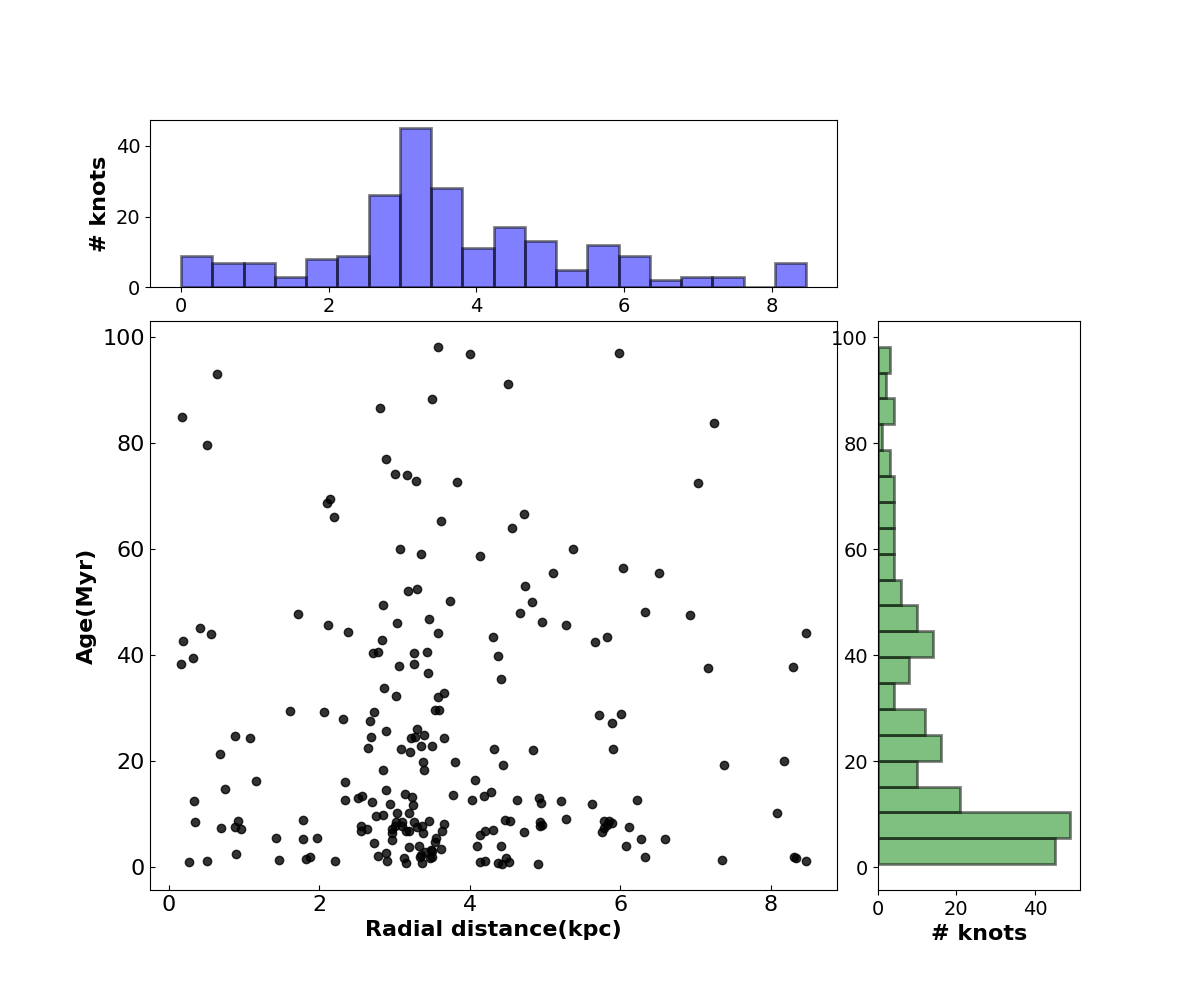}
     }
     
\hbox{
    \includegraphics[scale=0.32]{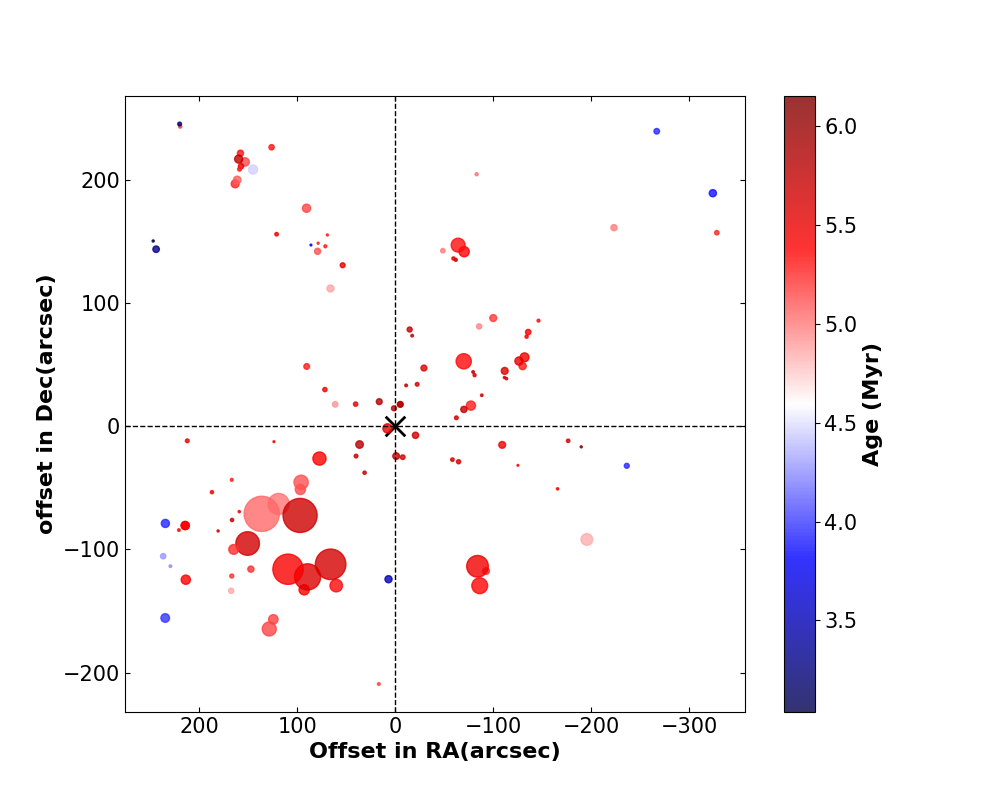}
     \includegraphics[scale=0.33]{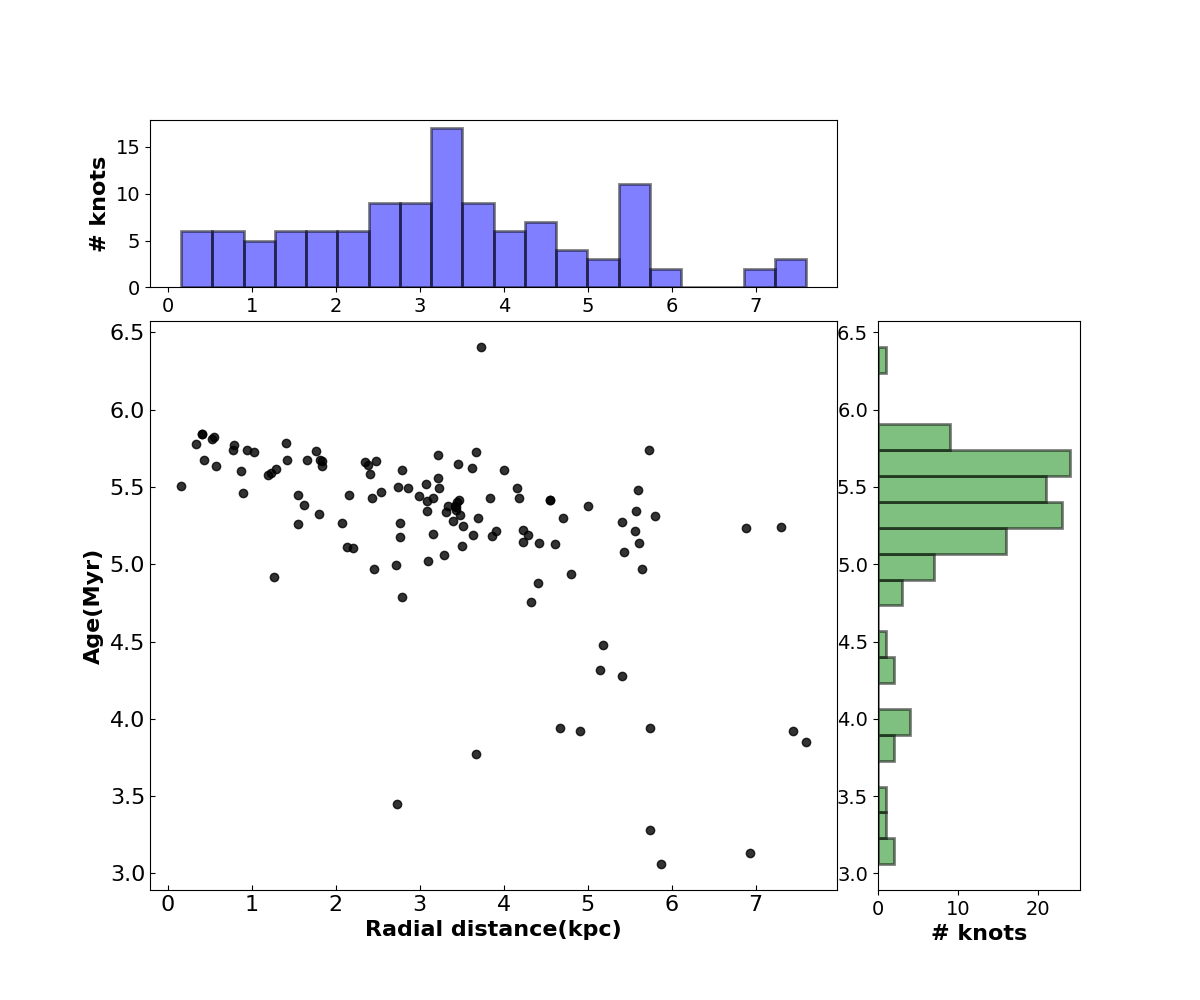}
     }
}
\caption{Distribution of age of the SF regions and
its spatial variation in F148W (top panel) and H$\alpha$ (bottom panel). In the left panel different sizes of the points correspond to the size of the SF regions and the cross is the position of the central AGN.
}
\label{figure-7}
\end{figure*}

To determine the age of the SF regions, we used 
the simple stellar evolutionary population model Starburst99 
\citep{1999ApJS..123....3L}. We generated synthetic stellar spectra using 
Starburst99 by adopting  Kroupa IMF,  total stellar mass of $10^9 M_\odot$, solar 
metallicity and ages upto 100 Myr. We convolved the generated UV spectra with the effective area of 
the respective UVIT filters to get the
fluxes in the corresponding UVIT filters as
\begin{equation}
F^{cal}(\lambda) = \frac{\int F(\lambda) EA(\lambda) d\lambda} {\int EA(\lambda) d\lambda}
\end{equation}
Here, EA($\lambda$) is the effective area of the filters.  $F^{cal}(\lambda)$ 
is the simulated fluxes, which were then converted to AB magnitudes
using the zero points given in \cite{2020AJ....159..158T}. These were then used
to generate the theoretical colours. The plot of the age against the theoretical
colour is given in Fig.~\ref{fig:figure-6}.  The observed FUV$-$NUV colours of 
the SF regions were compared with the theoretical colours to get 
the ages of the 
SF regions. The distribution of the ages of the SF regions
obtained from UV colour, as well as its spatial variation in NGC~4395 are shown 
in the top panel of Fig.~\ref{figure-7}.

\subsubsection{H$\alpha$ equivalent width}
For the observations carried out in H$\alpha$, we determined the age of the 
SF regions using H$\alpha$ equivalent width. Here too, we  used
the  Starburst99 model \citep{1999ApJS..123....3L} using the same input parameters as explained in Section 4.2.1. Starburst99 generates 
the expected equivalent width for different stellar populations at different time scales. We calculated the equivalent 
width from the calibrated H$\alpha$ narrow band and nearest broad R-band 
continuum images using the following relation \citep{1990PASP..102.1217W}
\begin{equation}
EW_{line}=\frac{[f_\lambda(N)-f_\lambda(B)]W(N)}{f_\lambda(B) -[f_\lambda(N)W(N)/W(B)]}
\label{eq:eqw}
\end{equation}
Here, $f_\lambda(N)$ and $f_\lambda(B)$ are the flux densities measured
in the narrow band H$\alpha$ and broad R-band filters. Similarly, 
W(N) and W(B) are the bandwidths of the H$\alpha$ line and broad R-band filters. 
The distribution of age determined from H$\alpha$ observations is 
shown in the bottom panel of Fig.~\ref{figure-7}. 

\subsection{Radial dependence of SFR and age}
The spatial distribution of the SF regions of varying ages can provide
information on the impact of the local environment and the central AGN can 
have on the
star formation characteristics of NGC~4395. To check for any such variation,
we calculated the distance (R) and the deprojected distance (R$^\prime$) of the
SF regions from the central AGN. The distance R$^\prime$ is calculated
as
\begin{equation}
R^\prime = \frac{R}{\sqrt{(cos^2(\theta) + sin^2(\theta)cos^2(i))}}
\end{equation}
where $\theta$ is given as
\begin{equation}
\theta = cos^{-1}\frac{\delta DEC}{\sqrt{\delta RA^2 + \delta DEC^2}}
\end{equation}
here, R is the observed distance from the center of NGC~4395, $i$ is the inclination of the galaxy and $\theta$ is the angle between the major axis of the galaxy and the radial distance to the SF regions \citep{1995AJ....109.2428M,2007MNRAS.381..943G}. 
The variation of the age of the SF regions
obtained from UV color, as a function of $R^\prime$ is shown in 
Fig.~\ref{figure-7}. The distribution is random with no indication of 
variation of the age of the SF regions with distance as shown in
 top panel of Fig.~\ref{figure-7}. In the bottom panel of Fig.~\ref{figure-7} is the variation of the age of the 
SF regions obtained from H$\alpha$ as a function of galactocentric distance. 
There is a trend of young SF regions occupying regions at increasing
distances from the central AGN. However, such a behaviour is not seen in UV. This different trend seen in the age versus radial distance in H$\alpha$ and UV  could be due to the different stellar populations probed by  the integrated UV and H$\alpha$ emission. Any age gradients that might be in place  during  the short time scales (as probed by H$\alpha$), possibly are obliterated on the time scales probed by UV (likely due to radial migration etc.). The complete catalogue of the 
SF regions identified in UV is given in Table \ref{table-3}.

\subsection{SFR at multiple wavelengths}
The SFRs of the SF regions identified in NGC~4395 were estimated using 
observations carried out in the UV and H$\alpha$ bands taking into account the differences in the instrumental resolutions in UV and H$\alpha$.
There are different calibration relations available in 
the literature \citep{1998ARA&A..36..189K,2007ApJS..173..267S,2017ApJ...847..136B} that pertain to observations at
different wavelengths. It would be ideal for getting the SFR of the SF regions
identified in F148W and H$\alpha$ at other wavelengths so that 
a comparison of the SFR estimates from different wavelengths could be 
made. But, for the regions identified in F148W and H$\alpha$
it is not possible to get the SFR at other longer wavelengths, because
of poorer resolution at longer wavelengths. 

\begin{figure}
    \centering
    \includegraphics[scale=0.3]{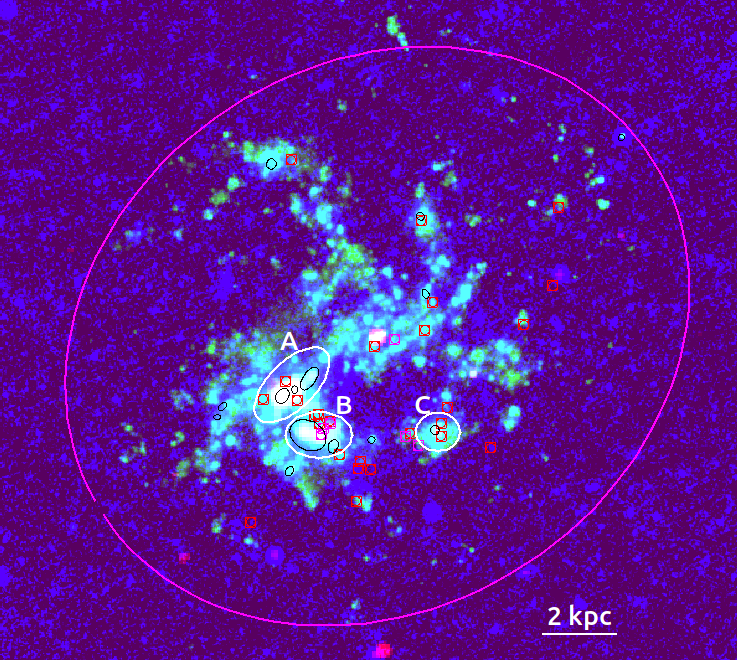}
    \caption{Composite image of NGC~4395 in 1.4 GHz continuum (red), GALEX FUV (green), and 24 $\mu$m MIPS (blue). The identified SF regions based on the 24 $\mu$m image are overlaid on the image by black ellipses. The white ellipses marked by A, B, C are the detected SF region having counterparts in 1.4 GHz. The big magenta ellipse is the aperture equivalent to D25 parameter in the optical U-band. The red small boxes are the positions of SNe remnants\citep{2013MNRAS.429..189L}.}
    \label{figure-8}
\end{figure}

An approach to compare the SFR estimated at different wavelengths is to
identify SF regions in images of different wavelengths at
similar resolution. To achieve this, we 
used the image in 24 $\mu$m from  
{\it Spitzer MIPS} and smoothed the images of 
NGC~4395 acquired at other wavelengths 
such as the UV and optical bands to the spatial resolution of 
24 $\mu$m image. This was done using the task HGEOM in the AIPS software \citep{1985daa..conf..195W}. We then identified SF regions in the 24 $\mu$m image
using the same procedure outlined in Section 3.1.
We identified a total of 14 SF regions. 
We show in Fig.~\ref{figure-8} the composite image of NGC~4395 with the identified
14 regions marked by black ellipses on it.
For these 14 SF regions, we calculated the SFR in FUV, NUV, $H\alpha$, and 24 $\mu$m. For UV and H$\alpha$ we used the calibration relations
given in Equations 8 and 9, while for 24 $\mu$m we 
used the following calibration relation 
\citep{2017ApJ...847..136B}. 

\begin{equation}
log(SFR_{24\mu m}) = [log[L_{24 \mu m}(erg.s^{-1})] - 40.93]/1.3 -1.26
\end{equation}

\begin{figure}
    \centering
    \includegraphics[scale=0.4]{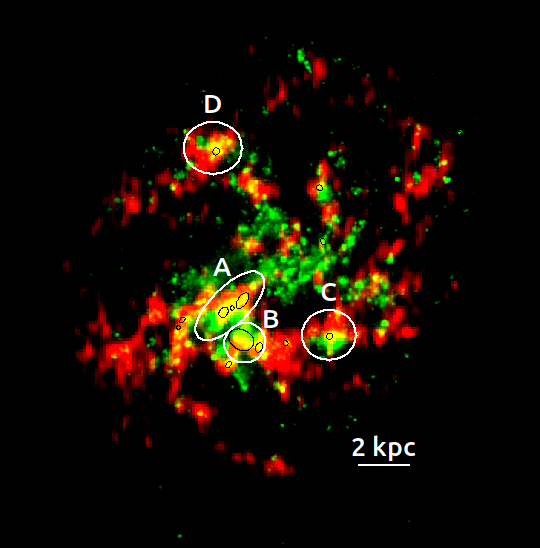}
    \caption{Composite image of NGC~4395 in H{\sc i} line image \citep{2008ASPC..396..267H} (red) and GALEX FUV (green). Black ellipses are the identified regions from 24 $\mu$m MIPS image. The white ellipses marked by A, B, C, and D are the SF complexes with strong UV and H{\sc i} emissions.}
    \label{fig:HI_fuv}
\end{figure}

The results obtained from different wavelengths for these 14 SF regions are 
given in Table \ref{table-4}. The values of SFR were estimated from the smoothed images, so that they have similar resolutions. Therefore, there will not be any effect of resolution on the estimated SFR (given in Table \ref{table-4}) for different filters. Out of these 14 detected SF regions in IR, 6 SF regions (3rd, 6th, 7th, 8th, 9th, and 10th SF regions in Table \ref{table-4}) have higher SFR in all wavelengths. In Fig.~\ref{figure-8} we have marked three star forming complexes (A, B, and C) by white ellipses containing 3, 2, and 1 SF regions, respectively detected based on 24 $\mu$m band. These three SF complexes are known to be associated with SNe remnants \citep{2005SerAJ.170..101V,2013MNRAS.429..189L}.
Using an aperture of 45 $\times$ 45 square arcsec  centered around the complexes  A, B, and C we calculated the number density of resolved SF regions detected in the high-resolution UV image from UVIT-$\it{AstroSat}$ and compared it with nearby 45 $\times$ 45 square arcsec SNe-free regions. In regions centered around the complexes A, B, and C the average number density of SF regions is 1.7$\times$10$^{-5}$ pc$^{-2}$ whereas in the regions devoid of any known SNe, it is 0.7$\times$10$^{-5}$ pc$^{-2}$. 
The enhanced flux density and higher number  of resolved SF regions in these three complexes (A, B, and C) could be thus attributed to positive feedback effects from SNe. 
As shown in Fig.~\ref{fig:HI_fuv} the SF regions associated with A, B, and C have strong H{\sc i} gas components, and also other SF region (13th in Table \ref{table-4}) marked as D has strong H{\sc i} emission based on the images from the WSRT \citep{2008ASPC..396..267H}.

\begin{table*}
\centering
\caption {The SFR in multiple wavelengths determined for the 14 SF regions
identified in the  MIPS 24 $\mu$m image. Here, a, b are respectively the semi-major and semi-minor axes of the SF regions, while PA is the position angle.}
\begin{tabular}{cccccccccc} \hline
No & RA & DEC & a  & b  & PA & \multicolumn{4}{c} {SFR(10$^{-3}$ M$_{\odot}$ yr$^{-1}$)}   \\ \cline{7-10}
   & (deg)      & (deg)      & arcsec & arcsec & (deg)   &  FUV              &  NUV                &   H$\alpha$         &  24 $\mu$m    \\ \hline
 1 & 186.488535 &  33.502065 &  6.06 &  4.24 &  57.5 & 0.152 $\pm$ 0.003 & 0.140 $\pm$ 0.002 &  0.717 $\pm$ 0.002 & 1.446 $\pm$ 0.333    \\
 2 &  186.455487 &  33.512476 & 4.37 & 4.09 & -14.7 & 0.070 $\pm$ 0.002 & 0.063 $\pm$0.001 & 0.999 $\pm$ 0.002 & 1.677 $\pm$ 0.359 \\
 3 &  186.429978 & 33.515758 &  5.53 & 5.16 & 88.3 & 0.672 $\pm$ 0.014 & 0.646 $\pm$ 0.007 & 1.894 $\pm$ 0.003 & 1.182 $\pm$ 0.301 \\
 4 & 186.517630 &  33.520106 &   4.16 & 3.10 & -2.6 & 0.023 $\pm$ 0.001 & 0.024 $\pm$ 0.001 & 0.070 $\pm$ 0.001 & 0.826 $\pm$ 0.252 \\
 5 & 186.515393 & 33.523667 &   5.97 &   3.67 & 42.76 & 0.071 $\pm$ 0.002 & 0.070 $\pm$ 0.001 &  0.179 $\pm$ 0.001 & 0.367 $\pm$ 0.168    \\
 6 & 186.491412 &  33.527326 &   9.88 &  7.319 &   52.35 & 2.9 $\pm$ 0.059 & 2.789 $\pm$ 0.031 & 13.493 $\pm$ 0.008 & 10.695 $\pm$ 0.906  \\
 7 & 186.486431 &  33.529412 &   4.51 &   3.76 & -89.7 & 0.359 $\pm$ 0.008 & 0.362 $\pm$ 0.004 & 3.107 $\pm$ 0.004 & 2.532 $\pm$ 0.441 \\
 8 & 186.480444 &  33.533089 &  15.74 &   7.71 & 56.1 & 0.664 $\pm$ 0.014 & 0.689 $\pm$ 0.008 & 3.373 $\pm$ 0.004 & 3.757 $\pm$ 0.537 \\
 9 & 186.470888 &  33.510386 &   8.42 &  5.99 &  71.9 & 0.296 $\pm$ 0.006 & 0.303 $\pm$ 0.003 & 1.643 $\pm$ 0.003 & 1.728 $\pm$ 0.364    \\
10 & 186.481081 &  33.514092 &  23.10 & 17.29 &  -30.1 & 4.178 $\pm$ 0.044 & 4.178 $\pm$ 0.044 & 12.080 $\pm$ 0.007 & 6.421 $\pm$ 0.702   \\
11 & 186.433656 &  33.561573 &   5.94 &  4.02 &  -66.94  & 0.135 $\pm$ 0.003 & 0.128 $\pm$ 0.002 & 1.064 $\pm$ 0.002 & 0.934 $\pm$ 0.268    \\
12 & 186.435697 &  33.587645 &   5.15 & 4.65 & -37.0  & 0.297 $\pm$ 0.006 & 0.270 $\pm$ 0.003 & 1.021 $\pm$ 0.002 &  0.909 $\pm$ 0.264    \\
13 & 186.495866 &  33.605167 &  6.37 &   5.72 & 64.5  & 0.206 $\pm$ 0.004 & 0.202 $\pm$ 0.002 & 0.862 $\pm$ 0.002 & 1.047 $\pm$ 0.283     \\
14 & 186.354740 &  33.614038 &  3.97 &   3.53 & 6.4  & 0.125 $\pm$ 0.002 & 0.170 $\pm$ 0.002 & 0.006 $\pm$ 0.001 & 1.510 $\pm$ 0.340     \\
   \\ \hline
\end{tabular}
\label{table-4} 
\end{table*}

\subsection{Global star formation rate}
In the earlier sections, we derived the SFR as well as the ages for many SF 
regions identified in NGC~4395 utilizing scaling relations.
Here, we aim to characterise the global SF 
nature of NGC~4395. For this, we decided to use two approaches using (a)
the existing calibrations and (b) broad band spectral energy
distribution (SED) modelling. Broad band SED modelling requires the brightness of the source to be evaluated at multiple wavelengths. For this we chose an 
aperture radius in the optical U-band such that the surface brightness drops 
to 25 mag/arcsec$^2$. The same aperture was then used to estimate
the brightness of NGC~4395 in different wavelengths, such as   
the GALEX FUV, NUV,  optical U, B, V, R, I from NOT, 2MASS J, H, and Ks, WISE W1, W2, W3, and 
MIPS 24 $\mu$m, 70 $\mu$m, and 160 $\mu$m bands.
A colour composite of NGC~4395 in UV, IR, and radio, with the aperture marked
on it is given in Fig.~\ref{figure-8}. The observed brightness of NGC~4395
at 16 different wavelengths was used to generate the broad band
SED and then modelled using  CIGALE (Code Investigating GALaxy
Emission; \citealt{2019A&A...622A.103B}) to derive different physical 
properties of NGC~4395. 

\begin{figure*}
    \centering
    \vspace*{-8.0cm}
    \includegraphics[scale=0.45]{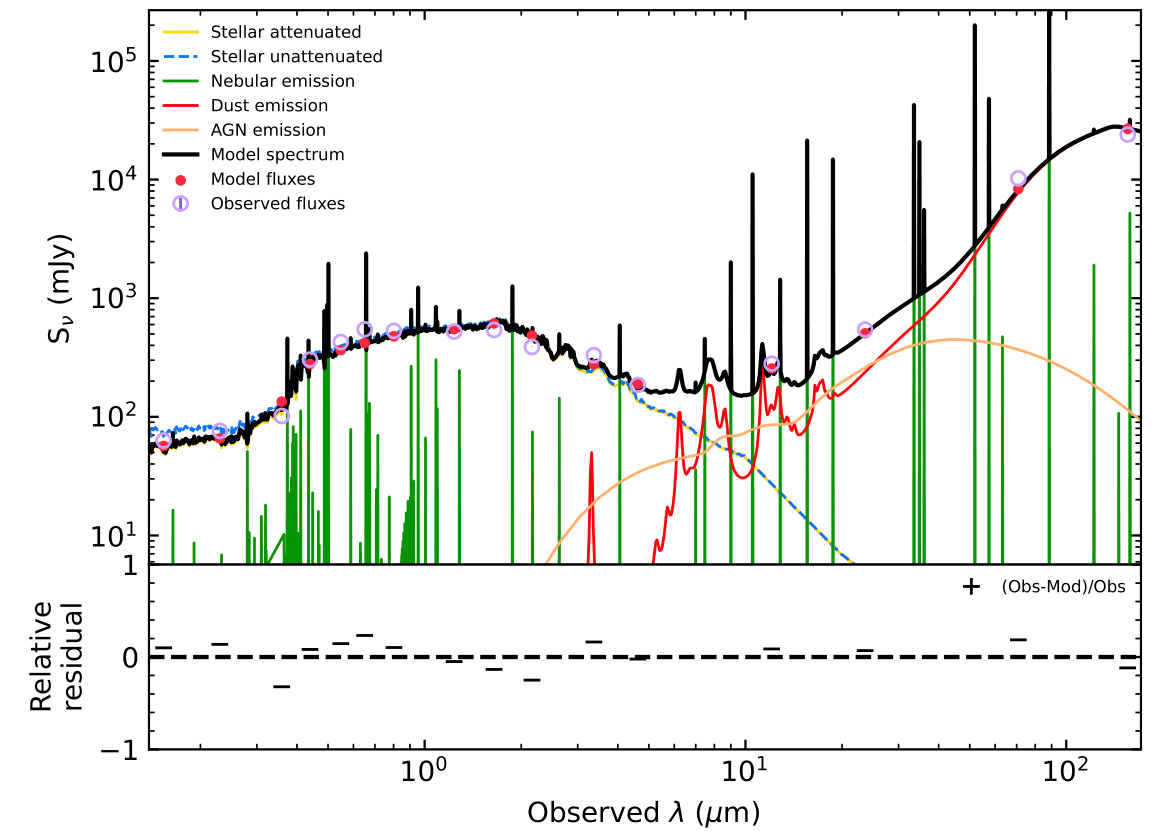}
    \caption{SED fitting using CIGALE code for the whole galaxy.}
    \label{figure-10}
\end{figure*}

We used CIGALE version 2022.0\footnote{https://cigale.lam.fr}. This
code works on the principle of energy balance and uses the Bayesian analysis 
method to derive various model parameters. To build the model, we used
the delayed star formation history with optional constant burst/quench module and having the form
$ SFR(t) \propto t \times exp(-t/\tau)$ for 0 $<$ t $<$ t$_0$. Here,
$t_0$ is the age of the onset of star formation and $\tau$ is the time
at which the SFR peaks. To model the stellar emission, we used
the \cite{2003MNRAS.344.1000B} single stellar population template with
Salpeter IMF. We used the \cite{2000ApJ...539..718C} model to take care of dust attenuation. The IR part of the SED because
of dust heated by stars was taken into account by the use of 
\cite{2014ApJ...780..172D, 2007ApJ...663..866D} model and
the presence of AGN was considered with the inclusion of 
SKIRTOR templates \citep{2016MNRAS.458.2288S}. The list of modules
and the  parameters used to build the SED of 
NGC~4395 are given in 
Table \ref{table-5}.  
The observed SED along with 
the model fit is given in Fig.~\ref{figure-10}. Some of the derived 
parameters from fits to the SED are given in Table \ref{table-6}. 
We also derived the global 
SFR in FUV, NUV, H$\alpha$, 24 $\mu$m, and 1.4 GHz using the scaling relations
given in Equations 8, 9, 14, and 15 \citep{2011ApJ...737...67M}.
\begin{equation}
log (SFR_{1.4 GHz})= log (L_{1.4}(erg. s^{-1}. Hz^{-1})) - 28.20
\end{equation}
It is found that while the 
SFR obtained in FUV, NUV, H$\alpha$, 1.4 GHz and SED fitting are nearly in agreement with each other as well as reported by \cite{2009ApJ...706..599L}, 
the SFR obtained in 24 $\mu$m is lower. 
A summary of these results is given in 
Table \ref{table-7}. As can be seen in Table \ref{table-7}, the 
SFR of about 0.5 M$_\odot$yr$^{-1}$ found from few tracers are in agreement 
with each other. However, from 24 $\mu$m IR emission we obtained a SFR of 0.05 M$_\odot$yr$^{-1}$ which is in agreement with the value of the SFR of 0.03 M$_\odot$yr$^{-1}$ found by \cite{2020AstBu..75..234S}. This value is lower than that found from other tracers such as FUV, NUV, H$\alpha$ and 1.4 GHz. The lower SFR at 24 $\mu$m is due to lower intrinsic luminosity at 24 $\mu$m than that predicted flux (predicted 24 $\mu$m luminosity $\sim$ 10 times the observed 24 $\mu$m luminosity) from H$\alpha$ luminosity by \cite{2009ApJ...703.1672K} for star-forming galaxies. The SFR calculated from the predicted luminosity at 24 $\mu$m is similar to that obtained from other tracers.

\begin{table*}
\caption {Input parameters for SED fitting with CIGALE}
\begin{tabular} {p{0.35\linewidth} p{0.55\linewidth}} \hline
 Parameter & Value \\ \hline
\textbf{Star Formation History  (sfhdelayedbq)}   &  \\
 e-folding time of the main stellar population &  200.0, 500.0, 700.0, 1000.0, 2000.0, 3000.0, 4000.0, 5000.0 Myr\\
  Age of main stellar population & 1500, 2000 Myr \\
 
 Age of the late burst/quench episode & 500 Myr \\
 Ratio of the SFR after$/$before burst/quench & 0.0,0.1,0.2,0.4,0.7,1.0,5.0,10.0,100.0 \\
 instance without any burst & 1.0 \\
 \hline
\textbf{ Stellar population synthesis (bc03)} & \\
 Initial mass fraction & Salpeter \\
  Metalicity & 0.008,0.02, 0.05 \\
 Separation between the young and the old stellar population & 10 Myr\\ \hline
 \textbf{Nebular emission (nebular)}    & \\ 
  Ionisation parameter & -2.0 \\
Electron density & 100 cm$^{-3}$\\

   \hline
\textbf{Dust attenuation (dustatt$\_$modified$\_$CF00)}  &  \\
 V-band attenuation in the interstellar medium & 0.05,0.10,0.15,0.2,0.5 \\
  AV\_ISM / (Av\_BC + Av\_ISM) & 0.50 \\
  Power law slope of the attenuation in the ISM & -0.7\\
  Power law slope of the attenuation in the birth clouds & -0.7 \\
 Filters for which the attenuation will be computed and added & FUV and NUV \\
 \hline
 \textbf{Dust emission (dl2014)}   & \\
Mass fraction of PAH & 1.12,2.5,3.19,5.26 \\
Minimum radiation field & 1.0,10,15,25 \\
Power law slope  & 1.5, 2.0 \\ 
Fraction illuminated from Umin to Umax &   0.001,0.01,0.1,0.5 \\
 \hline
 \textbf{AGN module (skirtor2016)}  & \\
Average edge-on optical depth at 9.7 micron & 3, 7 \\
power law exponent that sets radial gradient with polar angle & 1.0 \\
index that sets dust density gradient with polar angle & 1.0 \\
inclination, i.e. viewing angle, position of the instrument w.r.t. the AGN axis & 30, 70 \\
Disk spectrum & Skirtor spectrum \\
AGN fraction & 0.0, 0.05, 0.1, 0.2, 0.3, 0.4 \\
Wavelength range in microns were to compute the AGN fraction & 0/0 \\
Extinction law of the polar dust & SMC \\
E(B-V) for the extinction in the polar direction in magnitudes & 0.0, 0.03, 0.2, 0.4 \\
Temperature of the polar dust & 100.0 K \\
Emissivity index of the polar dust & 1.6 \\
 \hline
 
\end{tabular}
\label{table-5}
\end{table*}

\begin{table}
\caption {Parameters derived from broad band SED analysis}
\begin{tabular}{cc} \hline
Parameter    & Value  \\ \hline
  Stellar mass(M$_\odot$)  & 1.01 $\times$ 10$^9$ \\
SFR(M$_\odot$yr$^{-1}$)      & 0.21 $\pm$ 0.02  \\
SFR$_{100Myrs}$(M$_\odot$yr$^{-1}$) & 0.21 $\pm$ 0.02  \\
SFR$_{10Myrs}$(M$_\odot$yr$^{-1}$)   & 0.21 $\pm$ 0.02 \\
Attenuation(Av$\_${ISM}) & 0.1 \\
AGN fraction & 0.1 \\ \hline
\end{tabular}
\label{table-6}
\end{table}

\begin{table}
\caption {Summary of SFR obtained at different wavelengths. A value of 0.0 in the error of SFR means that the value is lesser than $1 \times 10^{-3}$. }
\begin{tabular}{cc} \hline
Tracer    & SFR (M$_{\odot}$yr$^{-1}$)  \\ \hline
FUV       & $0.47 \pm 0.00 $ \\
NUV       & $0.55 \pm 0.00 $  \\
H$\alpha$ & $ 0.32 \pm 0.00$ \\
24 $\mu$m  & 0.05 \\
1.4 GHz   & 0.52  \\
SED & 0.21 $\pm$ 0.02 \\ \hline
\end{tabular}
\label{table-7}
\end{table}

\section{Discussion}
Recent theoretical studies point to AGN feedback operating in dwarf galaxies \citep{2022MNRAS.516.2112K}, which is also supported by observations \citep{2018MNRAS.476..979P}. We have carried out a multiwavelength investigation on the star formation characteristics of the dwarf AGN 
NGC~4395 to understand star formation and the effect of AGN on the ISM of the host galaxy. Within 15 arcsec from the center of NGC~4395, we noticed three SF regions
having high SFR in UV. The high star formation in these regions in close proximity to the AGN could be because of positive feedback from AGN. From SED model fits we obtained an AGN fraction around 0.1 (see Table \ref{table-6}) which too points to the contribution of AGN to the observed properties. Alternatively, such enhanced star formation close to the center of the AGN can happen if gas can be funnelled into the center of the galaxy by the presence of bar and/or tidal interaction. However, in the case of NGC 4395, we do not have evidence of bar or tidal interaction. Using the stellar mass as returned by our CIGALE modelling of the galaxy and SFR obtained from FUV, we calculated the specific star formation rate (sSFR = SFR per unit mass). We found a value of sSFR = 4.64 $\times$ 10$^{-10}$ yr$^{-1}$. This is larger than that of M33 (sSFR= 1.13 $\times$ 10$^{-10}$ yr$^{-1}$;  \citealt{2009A&A...493..453V,2003MNRAS.342..199C})  and LMC (sSFR = 7.41 $\times$ 10$^{-11}$ yr$^{-1}$; \citealt{2009AJ....138.1243H,2002AJ....124.2639V}). These galaxies have stellar masses (M$_{\ast}|_{M33}$ = 4 $\times$  $10^{9}$ M$_{\odot}$; \citealt{2003MNRAS.342..199C} and M$_{\ast}|_{LMC}$= 2.7 $\times$ 10$^{9}$ M$_{\odot}$; \citealt{2002AJ....124.2639V}) similar to that of NGC~4395.
We noticed a gradual decrease in the age of the SF regions in H$\alpha$ from the center towards the outer regions; however, this is not seen in the SF regions identified from UV.

At 1.4 GHz we identified three SF complexes with enhanced radio emission marked as A, B, and C in Fig. \ref{figure-8}. These SF complexes encompass few resolved SF regions seen in IR and UV showing higher SFR. Also, the number density of resolved SF regions identified in high-resolution UV is higher in these regions than the surroundings. Massive stars in these SF regions can affect the ISM through stellar winds and SNe explosions. SNe explosions could have an impact on the ISM which could be either positive feedback via enhancement of star formation \citep{2014IAUS..296..265H} or negative feedback via inhibition of star formation \citep{2023arXiv230313574G}, however, the details are highly uncertain.

All the three SF complexes namely A, B, and C are known to be associated with SNe remnants \citep{2005SerAJ.170..101V,2013MNRAS.429..189L}. The high SFR and high number density of SF regions seen in SF complexes A, B, and C, could be due to positive feedback effects from SNe.
Alternatively, one would also expect higher SNe rates in areas of high star formation due to stellar feedback processes. However, we found many
regions that have a supernova remnant without any associated bright star forming regions (see Fig. \ref{figure-8}). It is probable that the high SFR seen in the SF complexes A, B and C is due to positive feedback from SNe. However, we note here that, just by the spatial coincidence of SNe and massive
star forming regions vis-a-vis isolated SNe not spatially coinciding with SF regions, it is difficult to conclude that SNe explosions could have caused
the high SFR via positive feedback. This is because, stars are known to migrate from their birth places, which is also supported theoretically \citep{2018MNRAS.481.1645M}. Therefore, one needs to account for the effects of radial migration before any conclusive claims on SNe
induced positive feedback effects could be made. However, the SF regions in complex A (regions 6,7 and 8 in Table \ref{table-4}), complex B (regions 9,10 in Table \ref{table-4}) and complex C (region 3 in Table \ref{table-4}) have higher SFR in H$\alpha$ and 24 $\mu$m (see Table \ref{table-4}) compared to other SF regions, which could be due to the compression of ISM during SNe explosions, arguing for SNe induced star formation, as in the case of NGC~2770 \citep{2020A&A...642A..84M}.

We found that cold neutral gas (H{$\sc$ i}) is extended throughout the spiral arms of the galaxy. However, there is no one to one correspondence between the SF regions traced by UV and H{$\sc$ i}, although the peaks of H{$\sc$ i} emission \citep{2008ASPC..396..267H} are associated with regions of high star formation and strong UV flux. To have a more detailed understanding of the relationship between UV and H{$\sc$ i} emission, H{$\sc$ i} observations with spatial resolution comparable to UV are needed. Systematic investigation of the star formation characteristics of a large sample of dwarf galaxies is needed to enhance our understanding of the complex feedback processes (supernova and/or AGN) operating in dwarf galaxies.

\section{Summary}
In this work, we have carried out UV and H$\alpha$ observations to study 
the spatially resolved SF regions of the host of the dwarf AGN, NGC~4395. 
In addition to the new observations in UV and H$\alpha$, we also
used archival data in UV, IR, and radio wavelengths. We have also studied the 
global star formation properties of NGC~4395  using UV, optical, IR, and radio 
data. The results of this work are summarized below.
\begin{enumerate}
\item Using UV data acquired from UVIT onboard $\it{AstroSat}$,  we identified
a total of 284 SF regions from the F148W image, extending up to a 
distance of 9 kpc.
\item Of the 284 SF regions identified in UV, 120 SF regions were also
identified in the H$\alpha$ continuum subtracted image. The detection 
of fewer SF regions in H$\alpha$ is attributed to the lower spatial resolution as well as the shallowness of the H$\alpha$ image relative to the UV one.
\item We found the SFR in F148W to lie between 
2.0 $\times 10^{-5}$ M$_\odot$yr$^{-1}$ and 1.5 $\times 10^{-2}$ M$_\odot$yr$^{-1}$ 
with a median value of 3.0 $\times 10^{-4}$ M$_\odot$yr$^{-1}$.
Using UV observations we found the 
age of SF regions to vary between 1 Myr and 98 Myr with a 
median value of 14 Myr.
\item In H$\alpha$ we found a median SFR of 1.7$\times 10^{-4}$ M$_\odot$yr$^{-1}$ with values between 7.2$\times 10^{-6} $ M$_\odot$yr$^{-1}$ and 2.7 $\times 10^{-2}$ M$_\odot$yr$^{-1}$. For the SF regions in  H$\alpha$, 
we found a median age of 5 Myr with values ranging from 3 Myr to 
6 Myr.
\item We did not find any noticeable gradual variation of the ages or 
$\Sigma_{SFR}$ of the SF regions from the center to the outer regions of the 
galaxy in UV.
\item For the ages of the SF regions determined from H$\alpha$ equivalent 
width we found indication of a gradual decrease in their age from 
the center of the source outwards. This might be attributed to the intense
SF regions seen in H$\alpha$ in the spiral arms of NGC~4395.  
\item On inspection of the spatial distribution of $\Sigma_{SFR}$ in 
UV (see Fig.~\ref{figure-5}) we found  three SF regions near the  AGN having 
high $\Sigma_{SFR}$.  One out of the three SF regions in UV is also found to
have high  $\Sigma_{SFR}$ in H$\alpha$ (lower left of Fig.~\ref{figure-4}) and
a younger age (lower left of Fig.~\ref{figure-7}). This could possibly hint for a positive feedback from AGN. We, however, note that further observations are needed to confirm this.
\item We identified 14 common SF regions in UV to IR bands. Out of these 14, 7 SF knots  have associated H{\sc i} emission.
\item We found the global SFR of about 0.5 M$_\odot$yr$^{-1}$ in UV. This is consistent with the SFR determined from other tracers such as the H$\alpha$ and the 1.4 GHz radio continuum. We also calculated the sSFR of 4.64 $\times$ 10$^{-10}$ yr$^{-1}$ in NGC 4395, which is larger than the sSFR of other dwarf galaxies such as M33 and LMC that have stellar masses similar to NGC 4395.
\item At 1.4 GHz, we found few complexes having enhanced radio emission. These complexes contain larger number of SF regions, with majority of them having higher SFR. These complexes are known to host supernova remnants. The SF regions in these complexes have higher SFR in H$\alpha$  and 24 $\mu$m, compared to other SF regions, arguing to supernova induced star formation. 

\end{enumerate}

\section*{Acknowledgements}
The authors thank the anonymous referee for his/her critical comments on the manuscript, leading to its improvement. This publication uses the data from the AstroSat mission of the Indian Space Research Organization (ISRO), archived at the Indian Space Science Data Center (ISSDC). This publication uses UVIT data processed by the  payload operations center at IIA. The UVIT is built in collaboration between IIA, IUCAA, TIFR, ISRO, and CSA. We thank the staff of IAO,  Hanle, and CREST, Hoskote that made the observations possible. The facilities at IAO and CREST are operated by the Indian Institute of Astrophysics, Bangalore. This work has made use of the NASA Astrophysics Data System\footnote{https://ui.adsabs.harvard.edu/} (ADS) and the NASA/IPAC extragalactic database\footnote{https://ned.ipac.caltech.edu} (NED). This work has made use of data from the European Space Agency (ESA) mission {\it Gaia} (\url{https://www.cosmos.esa.int/gaia}), processed by the {\it Gaia} Data Processing and Analysis Consortium (DPAC, \url{https://www.cosmos.esa.int/web/gaia/dpac/consortium}). Funding for the DPAC has been provided by national institutions, in particular, the institutions participating in the {\it Gaia} Multilateral Agreement. The authors acknowledge Dr. Tom Oosterloo for sharing the spectral data cube of the H{\sc i} data. A few of the authors thank the Alexander von Humboldt Foundation, Germany, for the award of the Group Linkage long-term research program.

\software{IRAF \citep{1986SPIE..627..733T}, SExtractor \citep{1996A&AS..117..393B}, CIGALE \citep{2019A&A...622A.103B}, Photutils \citep{larry_bradley_2020_4044744}, AIPS \citep{1985daa..conf..195W}}



\bibliography{ref}{}
\bibliographystyle{aasjournal} 


\label{lastpage}
\end{document}